\documentclass[preprint,12pt]{elsarticle}

\graphicspath{{figures/}}
\usepackage{amstext}

\usepackage{hyperref}

\usepackage{amssymb}

\usepackage{subfigure}

\usepackage{lineno}

\usepackage{xcolor}
\usepackage{xifthen}
\newcommand{\cmnt}[2][]{
    {\ifthenelse
            {\isempty{#1}}
            {\color{red}{#2}}
            {\color{red}{#2}\footnote{\color{red}{#1}}}
        }
}
\newcommand{\cmntd}[2][]{
    {\ifthenelse
            {\isempty{#1}}
            {\color{black}{#2}}
            {\color{black}{#2}$\!\!\!\!\!$}
        }
}

\definecolor{color}{rgb}{1,0,1}

\newcommand{\cfnd}[1]{{}}

\newcommand{\editsf}[1]{{\color{black}{#1}}}

\usepackage{scalerel}
\newcommand{\chr}{{\Theta_{\scaleto{C/H}{4pt}}}}
\newcommand{\chrl}{{\theta_{\scaleto{C/H}{4pt}}}}
\newcommand{\chrnorm}{{\theta^*_{\scaleto{C/H}{4pt}}}}
\usepackage{pifont}
\usepackage{textcomp}
\newcommand{\Nc}{{N_{C}}}
\newcommand{\Nh}{{N_{H}}}
\newcommand{\nc}{{n_{C}}}
\newcommand{\nh}{{n_{H}}}
\newcommand{\Nfive}{{N_5}}
\newcommand{\Nsix}{{N_6}}
\newcommand{\Nseven}{{N_7}}
\newcommand{\Nring}{{N_{\scriptscriptstyle\bigcirc}}}
\newcommand{\Ncring}{{N_{\copyright}}}
\newcommand{\Ncnonring}{{N_{\mbox{\textcent}}}}
\newcommand{\NcringP}{{N^{\prime\prime}_{\copyright}}}
\newcommand{\NcnonringP}{{N^{\prime\prime}_{\mbox{\textcent}}}}
\newcommand{\ncring}{{n_{\copyright}}}
\newcommand{\ncnonring}{{n_{\mbox{\textcent}}}}
\usepackage[normalem]{ulem}
\usepackage{nicefrac}

\usepackage{eqnarray}
\usepackage{titlesec}
\begin{document}

\begin{frontmatter}

	\title{Internal Structure of Incipient Soot from Acetylene Pyrolysis obtained via Molecular Dynamics Simulations
	}

	\author[inst1]{Khaled Mosharraf Mukut}

	\affiliation[inst1]{organization={Department of Mechanical Engineering},
		addressline={ Marquette University},
		city={Milwaukee},
		state={Wisconsin},
		postcode={53233},
		country={USA}}

	\author[inst2]{Anindya Ganguly}
	\author[inst2]{Eirini Goudeli}

	\affiliation[inst2]{organization={Department of Chemical Engineering},
		addressline={University of Melbourne},
		state={Victoria},
		country={Australia}}

		\author[inst3,inst4]{Georgios A. Kelesidis}

		\affiliation[inst3]{organization={Nanoscience and Advanced Materials Center (NAMC), Environmental and Occupational Health Science Institute, School of Public Health, Rutgers, The State University of New Jersey},
			addressline={170 Frelinghuysen Road},
			city={Piscataway},
			state={New Jersey, 08854},
			country={U.S.A.}}
	
		\affiliation[inst4]{organization={Particle Technology Laboratory, Institute of Process Engineering, Department of Mechanical and Process Engineering},
			addressline={ETH Z$\ddot{u}$rich},
			city={Sonneggstrasse 3},
			state={CH-8092 Z$\ddot{u}$rich},
			country={Switzerland}}
	\author[inst1]{Somesh P. Roy\corref{cor1}}
    \cortext[cor1]{Corresponding author: somesh.roy@marquette.edu}
	\begin{abstract}

	A series of reactive molecular dynamics simulations is used to study the internal structure of incipient soot particles obtained from acetylene pyrolysis. The simulations were performed using ReaxFF potential at four different temperatures. The resulting soot particles are cataloged and analyzed to obtain statistics of their mass, volume, density, C/H ratio, number of cyclic structures, and other features. A total of 3324 incipient soot particles were analyzed in this study. \editsf{Based on their structural characteristics}, the incipient soot particles are classified into two classes, referred to as type~1 and type~2 incipient soot particles in this work. The radial distribution of density, \editsf{cyclic (5-, 6-, or 7-member rings)} structures and C/H ratio inside the particles revealed a clear difference in the internal structure between type~1 and type~2 particles. These classes were further found to be well represented by the size of the particles with smaller particles in type~1 and larger particles in type~2. The radial distributions of ring structures, density, and C/H ratio indicated the presence of a dense core region in type~2 particles, whereas no clear evidence of the presence of a core was found in type~1 particles. \editsf{In type 2 incipient soot particles, the boundary between the core and shell was found to be around 50\%--60\% of the particle radius of gyration}.

	\end{abstract}


\end{frontmatter}


\section{Introduction} \label{s:intro}

Soot is a harmful carbonaceous nanoparticle
generated during combustion of hydrocarbon fuels. Soot, also known as black carbon, can cause serious health
issues \cite{ Jacobsen2008Jul, Jerrett2013Aug} and acts as a major forcing factor in climate change\cite{Hansen2004Jan,Bond2013Jun}. The exact mechanism of the formation of soot particulates from gaseous precursors is still unknown due to the complex chemical nature of the hydrocarbon \editsf{reaction network} and time- and length-scale of the soot formation processes. According to the present understanding, soot formation occurs by a series of complex physicochemical
events such as the formation of gas-phase soot precursors including, but not limited to, polycyclic aromatic hydrocarbons (PAHs), nucleation of incipient soot particles, growth and maturation of incipient soot particles due \editsf{to} surface reactions, \editsf{aggregation by} coagulation or coalescence, and decay of the particles by fragmentation and oxidation~\cite{Irimiea2019Apr,Appel2000Apr, Wang2011Jan,Rigopoulos2019,Li2023}.
The inception of soot particles is arguably the least understood phenomenon among these processes and the exact chemical \editsf{reaction} pathways of soot inception are not completely known yet. Researchers agree that soot formation starts with forming small gas-phase precursor molecules such as acetylene which leads to PAHs like benzene, pyrene, and coronene~\cite{Dobbins1994, Balthasar2003May, Wang2016Feb}. The freshly formed PAHs then combine to form the solid or liquid-like incipient soot particles~\cite{Michelsen2020Oct,VanderWal1998Mar,Zhao2007Jan}. \editsf{These} particles then start to grow \editsf{by} surface \editsf{reactions} and coalescence to form larger soot particles \cite{Michelsen2020Oct,Mao2017Sep,Chen2023,Sabbah2010Oct,Sanchez2012Aug}.

Due to the complexity and scale of incipient \editsf{soot} particles, their exact internal structures are not very well characterized yet. Recent studies have shown young soot particles tend to have a condensed core of ring-like structures surrounded by a shell of highly stacked large molecules while surrounded by a shell of less stacked smaller molecules~\cite{Botero2019Jan}. As these incipient particles mature, their internal structures evolve, which in turn affects their physical and chemical properties. 

There have been some recent breakthroughs in the experimental exploration of the internal structure of incipient soot. For example, Chang et al.~\cite{Chang2020Jun} employed high-resolution transmission electron microscopy (HRTEM) and scanning electron microscopy (SEM)  to investigate the structural evolution and fragmentation of coal-derived soot and carbon black particles under high-temperature air oxidation conditions. They also explored the onset of micropores and the internal graphitic microcrystals using X-ray diffraction (XRD) and Raman spectra. Morajkar et al. \cite{Morajkar2020Nov} utilized HRTEM, XRD, Raman spectroscopy, and inductively coupled plasma mass spectrometry (ICPMS) to examine the transmission of trace metals from biodiesel fuels to soot particles and the nanostructural irregularities of the soot. In their study, Gleason et al.~\cite{Gleason2021Jan} indicated that the formation of soot nuclei in an ethylene/nitrogen flame can be attributed exclusively to aromatic compounds comprising one or two rings. Carbone et al.~\cite{Carbone2017Jul} conducted a comprehensive investigation of soot inception in a laminar premixed ethylene flame and found that soot particles undergo an aging transformation from being nearly transparent in the visible spectrum to a more graphitic-like composition. 
Using low-fluence laser desorption ionization (LDI) in conjunction with HRTEM, Jacobson et al. \cite{Jacobson2020Mar}, investigated the molecular composition of soot particles to determine the PAH concentration in soot particles.  The capabilities of atomic force microscopy (AFM) were exploited by Barone et. al.~\cite{Barone2003Jan} to calculate particle size distribution functions under different sampling conditions.  In their study, Schulz et al. \cite{Schulz2019Jan} conducted an investigation into the initial phases of soot formation using AFM and observed the presence of multiple aromatic compounds, some of which displayed noticeable aliphatic side chains. Commodo et al.~\cite{Commodo2017Jul} investigated the initial phases of soot formation using X-ray, ultraviolet photoemission spectroscopy (UPS), UV-visible, and Raman spectroscopy to show the coexistence of $sp^3$ carbon and a more advanced graphitic structure, which exhibits a slightly larger aromatic island, a reduced band gap, and an increased density of states. In another study, \editsf{Commodo et al.}~\cite{Commodo2019Jul} identified a noteworthy occurrence of aliphatic pentagonal rings in the early stages of soot formation, particularly in close proximity to the outer region of aromatic soot molecules, and it has been suggested that the elimination of hydrogen from these molecules can result in the creation of resonantly stabilized $\pi$-radicals \cite{Parker2011Feb}. This phenomenon has also been theorized by Johansson et al. \cite{Johansson2018Sep}, Gentile et al. \cite{Gentile2020Nov}, and Rundel et al. \cite{Rundel2022Sep}.

Even with such recent advancement in experimental findings, there is still a lot of unknowns about the internal structure of soot. The limitations of experimental methods can be compensated and complemented by first-principle modeling such as molecular dynamics. With the development of high-performance computational resources, reactive molecular dynamics (RMD) simulation has become more affordable for studying complex reactive networks.  For soot-relevant RMD studies, the reactive force field (ReaxFF) potential developed by van Duin et al. \cite{vanDuin2001Oct} for carbon, hydrogen, and oxygen chemistry (CHO-parameters \cite{Ashraf2017Feb,Chenoweth2008Feb}) is a popular choice. The ReaxFF potential can capture the physicochemical evolution of hydrocarbon systems in an extensive range of temperatures and pressures. It is based on the bond order between different atoms, which carry information related to bond breakage and formation.
In recent years, RMD simulations have been used to investigate soot nucleation by pyrene dimerization \cite{Schuetz2002Jan}, to shed light into the nucleation and growth of incipient soot \editsf{from} PAHs, such as naphthalene, pyrene, coronene, ovalene and circumcoronene \cite{Mao2017Sep}, to explore the initial mechanism of soot nanoparticle formation \cite{Han2017Aug} and to examine the effect of oxygenated additives on \editsf{the reduction of} diesel soot emissions \cite{Chen2020Sep}.

Since RMD simulation provides detailed structural information at the atomic scale, it can be an excellent tool for analyzing the internal structure of incipient soot particles. For example, recently Pascazio et al. \cite{Pascazio2020Sep} looked into the internal structure and the mechanical properties of incipient soot particles using RMD simulation and quantified the amount of cross-linking in the core and shell region of developing and mature soot particles. Mature soot primary particles exhibit a distinct core-shell structure with a disorderly condensed core of ring-like structures surrounded by a shell of chain-like structures~\cite{Botero2019Jan,Ishiguro1997Jan}.

 Process temperature plays an important role in the development and \editsf{aging} of soot particles. For example, in a recent study, Pathak et. al.~\cite{Pathak2022Apr} studied graphitization induced structural transformation of candle soot at different temperatures and found that increasing the temperature increases the rate of graphitization that leads to more spherical and mature soot and weakening of the correlation between graphitic nanostructure and surface functional groups (SFGs). SFGs have been found to be connected to the characteristics of soot aggregates, including the fractal dimension~\cite{Liu2017Mar}.  Since soot morphology, maturity and reactivity are expected to be influenced by temperature during acetylene pyrolysis \cite{Ruiz2007Influence}, it is important to study the internal structure of incipient soot particles at different temperatures.

In the present study, a series of isothermal RMD simulations using the ReaxFF potential is conducted mimicking acetylene pyrolysis at different temperatures (1350, 1500, 1650, and 1800K). A variety of physicochemical features of these RMD-generated soot particles are then analyzed to shed light on different types of incipient soot particles and to characterize the internal structure of these particles obtained from RMD simulations.

\section{Numerical Methodology} \label{s:numerics}
\subsection{Simulation configurations}
Following the methodology described in~\cite{Sharma2021Aug},  1000 acetylene molecules are randomly placed in a cubic domain ($75 \text{\AA} \times 75\text{\AA} \times 75 \text{\AA}$) at four different temperatures, i.e., 1350, 1500, 1650, and 1800~K. The temperatures are chosen to capture soot particles from various thermally activated systems. For
statistical significance, \editsf{at each temperature,} simulations are performed for \editsf{at least} five times with different initial \editsf{configurations}. In total, 24 RMD simulations were performed for four different temperatures. The RMD simulations are \editsf{performed} using the Large-scale Atomic/Molecular Massively Parallel Simulator (LAMMPS) \cite{Thompson2022Feb} software. ReaxFF potential for hydrocarbons \cite{vanDuin2001Oct, Castro-Marcano2012Mar} is used to capture the chemical changes (bond breakage and formation) \editsf{due to reactive} molecular collisions \editsf{during} acetylene \editsf{pyrolysis}. The bond length between individual atoms is calculated at each timestep (0.25 fs) based on the changes in the chemical environment to describe bond cleavage and formation accurately \cite{Chenoweth2008Feb}. This helps the model capture the chemical reactions leading to radical formation during soot nucleation. Periodic boundary
conditions are assumed in all three dimensions. The coordinates of each atom are calculated and updated using
the velocity-Verlet algorithm \cite{Swope1982Jan} in conjunction with the Nose-Hoover thermostat \cite{Evans1985Oct}. A constant number, volume, and temperature (NVT) ensemble strategy is used to run each simulation up to 10~ns. The simulation results are probed every \editsf{0.05}~ns and the clusters of hydrocarbons that resembles primary soot particles are isolated, \editsf{tabulated}, and analyzed. \editsf{Each of these extracted clusters has at least 20 carbon atoms and at least one 5-, 6-, or 7-membered ring structure following an earlier study by Mukut et al.\cite{Mukut2021May}.} Features such as surface area and volume of primary particles are calculated using MSMS software developed by Sanner \cite{Sanner1996Mar} and other physicochemical
characteristics are analyzed mostly using MAFIA-MD \cite{Mukut2022Jul}. The open visualization tool (OVITO)
\cite{Stukowski2009Dec} is used for visualization of the molecular clusters.

\subsection{Workflow}
The workflow in this study can be summarized as: 
\begin{enumerate}
	\item Conduct RMD simulations at different temperature with \editsf{various initial configurations.} and extract incipient soot particles from the trajectory results.
	\item Calculate chemical and morphological characteristics such as the number of
	      atoms, C/H ratio, the radius of gyration, atomic fractal dimension, density,
	      surface area, and volume.
	\item Classify the soot particles based on \textit{all} the calculated features using machine
	      learning techniques such as k-means clustering \cite{Lloyd1982Mar} and
	      t-distributed stochastic neighbor embedding (t-SNE) \cite{vanderMaaten2008}.
	\item Investigate the internal distribution of several relevant features such as
	      distribution of cyclic/non-cyclic molecules, C/H ratio, density etc. and find identifiable
	      patterns in the distribution.
\end{enumerate}

\subsection{Extraction of physicochemical properties} \label{s:physicochemical}
From the RMD simulations, we extract the coordinates of each atom present in the simulation box at regular time intervals via the trajectory file. Each individual timestep is investigated separately by analyzing the atom coordinates within the entire simulation domain, which contains both large molecular clusters and small molecules. The large soot-like molecular clusters are identified as the ones that have more than 20 carbon atoms and have at least one 5-, 6-, or 7-member ring~\cite{Mukut2021May}. It is noted that in our case, the smallest such cluster was found to have 65 carbon atoms. These clusters are isolated using the cluster analysis tool from the OVITO
Python module \cite{Stukowski2009Dec} implemented in a unified Python script developed inhouse. Then the isolated clusters are analyzed individually to calculate their physical, morphological and chemical attributes. Some attributes are obtained trivially from the trajectory files, e.g., number of atoms ($N$), carbon to hydrogen ratio ($\chr$), mass ($M_p$), and molar mass ($M$). Some other attributes like the radius of gyration ($R_g$), atomic fractal dimension ($D_f$), and density ($\rho$) are extracted by simple algebraic and geometric analysis or by using empirical correlations proposed in the literature and listed in~\ref{app_eqn}. The volume and the surface area of incipient particles are calculated using MSMS software~\cite{Sanner1996Mar} using a probe radius of 1.5~\AA.

The identification and analysis of 5-~/6-~/7-member ring structures are done using MAFIA-MD~\cite{Mukut2022Jul}. MAFIA-MD can analyze RMD trajectory files to identify cyclic/ring structures in an atomic cluster. Not all cyclic structures identified are necessarily aromatic. As discussed in~\cite{Mukut2022Jul}, it is difficult to exactly confirm which cyclic structures are aromatic as the information about aromaticity requires some approximations regarding the bond order of aromatic bonds and establishment of planarity. To remove any confusion, therefore, we used the terms ``ring'' or ``cyclic'' in this work instead of aromatic when discussing these internal structures in the soot clusters. The numbers of 5-, 6-, and 7-member rings are denoted as $\Nfive$, $\Nsix$, $\Nseven$, respectively, and the total number of rings is denoted as $\Nring$. Similarly, the number of carbons in rings is denoted as $\Ncring$ and the number of non-cyclic carbons in a particle is denoted as $\Ncnonring$.

A sample of two particles and a list of their respective properties are provided in~\ref{app_sample}. 
This entire set of features is used in the classification of particles as discussed in Sec.~\ref{ss:classification}.
It must be noted here that while all the above-mentioned properties were evaluated for each particle, this article only focuses on the internal structure of the particles, which are characterized as discussed in Sec.~\ref{s:internal-distr}. Therefore, beyond their use in the classification of particles, the detailed analysis of physical and morphological features such as volume, surface area, radius of gyration, and atomic fractal dimension are not the focus of this work.

\subsection{Characterization of internal structure} \label{s:internal-distr}
\begin{figure} [!htbp]
	\centering
	\includegraphics[width=0.7\linewidth]{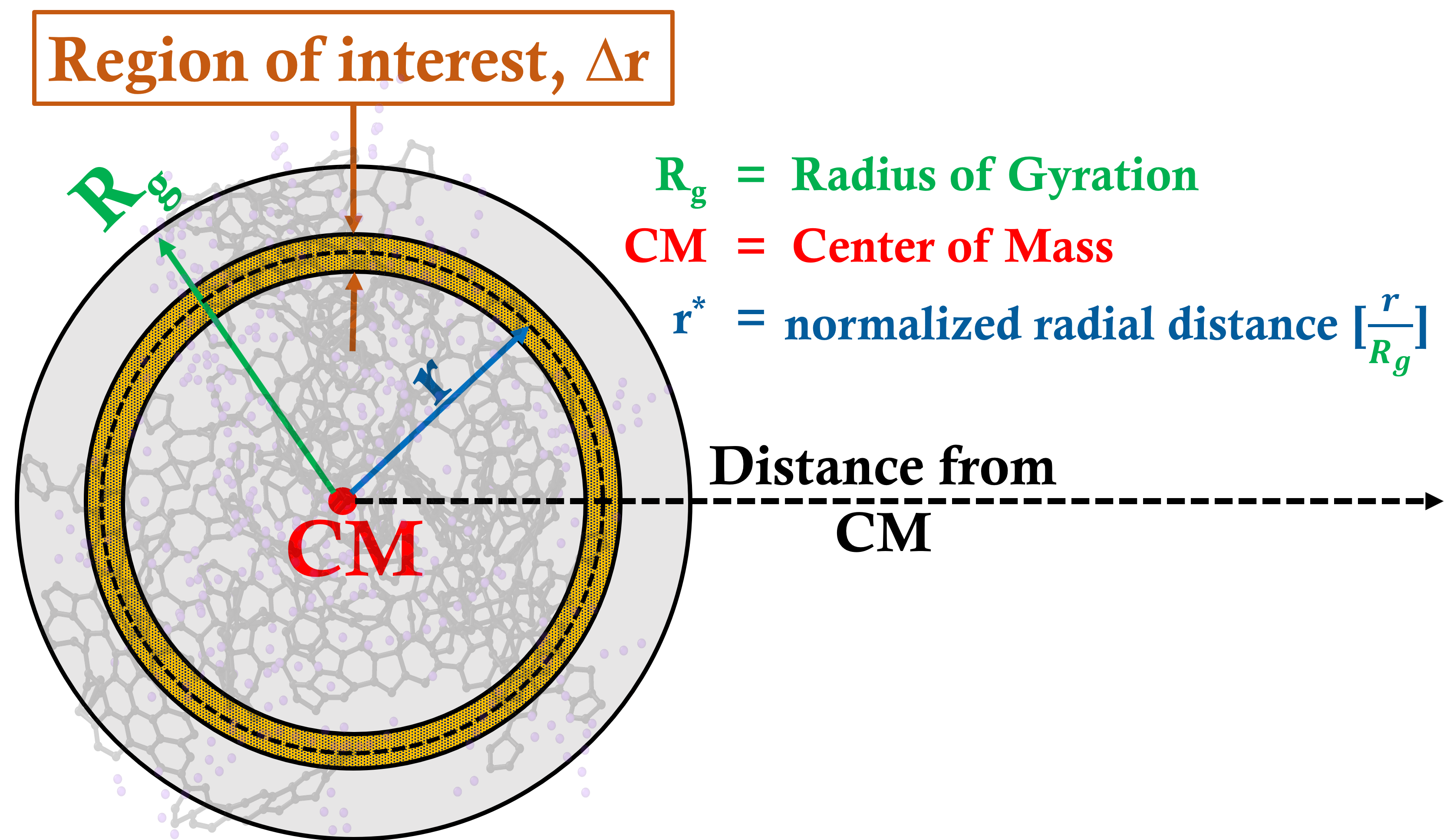}
	\caption{Schematic representation of the calculation of radial distribution of internal features in the soot particle.}
	\label{f:schematic}
\end{figure}

We analyzed the internal structure of soot particles via the radial distribution of carbon atoms, C/H ratio, and density inside the particle. In order to compare different-sized particles on the same scale, we first normalized the radius of particles by scaling each particle by its radius of gyration ($R_g$). Then each particle is divided in equal number of radial bins. Each radial bin creates a spherical shell or strip as shown by the shaded yellow region of interest in Fig.~\ref{f:schematic}. We calculate various internal features in these spherical strips and present them as a function of the normalized radial distance from the center of mass  of each spherical strip ($r^*=\nicefrac{r}{R_g}$).

For example, let's consider a strip of width $\Delta r$, whose midplane is distance $r$  away from the center of mass. The number of cyclic and non-cyclic carbon atoms in this strip (i.e., within a radial distance of $r\pm\nicefrac{\Delta r}{2}$) are counted and represented as \editsf{a function of} the normalized radius of the midplane ($r^*=\nicefrac{r}{R_g}$), as $\ncring(r^*)$ and $\ncnonring(r^*)$, respectively.
The radial distribution of cyclic and non-cyclic carbon per unit area (indicated by~$^{\prime\prime}$) at a normalized distance $r^*$ are then, respectively:
\begin{equation}
	\NcringP(r^*) = \frac{\ncring(r^*)}{4\pi (r^*\times R_g)^2};\qquad\NcnonringP(r^*) = \frac{\ncnonring(r^*)}{4\pi (r^*\times R_g)^2}\label{e:ring-surf-density}
\end{equation}

In a similar manner, the radial distribution of C/H ratio is also analyzed for each particle. The C/H ratio of the entire particle is calculated as $\chr=\nicefrac{\Nc}{\Nh}$, where $\Nc$ and $\Nh$ are number of carbon and hydrogen atoms in the entire particle, \editsf{respectively}. This C/H ratio is termed as the \textit{particle} C/H ratio ($\chr$) to differentiate from the \textit{local} C/H ratio ($\chrl$), which is calculated using the number of carbon and hydrogen atoms in the spherical strips as shown in Fig.~\ref{f:schematic}. The local C/H ratio ($\chrl$) is \editsf{determined} by calculating the number of carbon ($\nc(r^*)$) and hydrogen ($\nh(r^*)$) atoms in a spherical strip with the midplane at a normalized distance $r^*$ from the center of mass ($\chrl(r^*)=\nicefrac{\nc(r^*)}{\nh(r^*)}$). Finally, the local C/H ratio is normalized by the corresponding particle C/H ratio
\begin{equation}
	\chrnorm(r^*)=\frac{\chrl(r^*)}{\chr}\label{e:norm-chratio}
\end{equation}

Similarly, the radial distribution of local density is also evaluated by dividing the simulated density of the thin spherical strip using Eqn.~\ref{e:actual-denstiy} for the strip (referred as \textit{local} density, $\varrho(r^*)$) by the simulated density of the particle ($\rho_{s}$) as
\begin{equation}
	\rho^*_{s}(r^*) = \frac{\varrho(r^*)}{\rho_{s}}\label{e:norm-density}
\end{equation}

\section{Results and Discussion} \label{s:resultDiscussion}

\subsection{Formation of incipient soot particles in RMD} \label{ss:incipientSoot}
\begin{figure} [!htbp]
	\centering
	\includegraphics[width=0.9\linewidth]{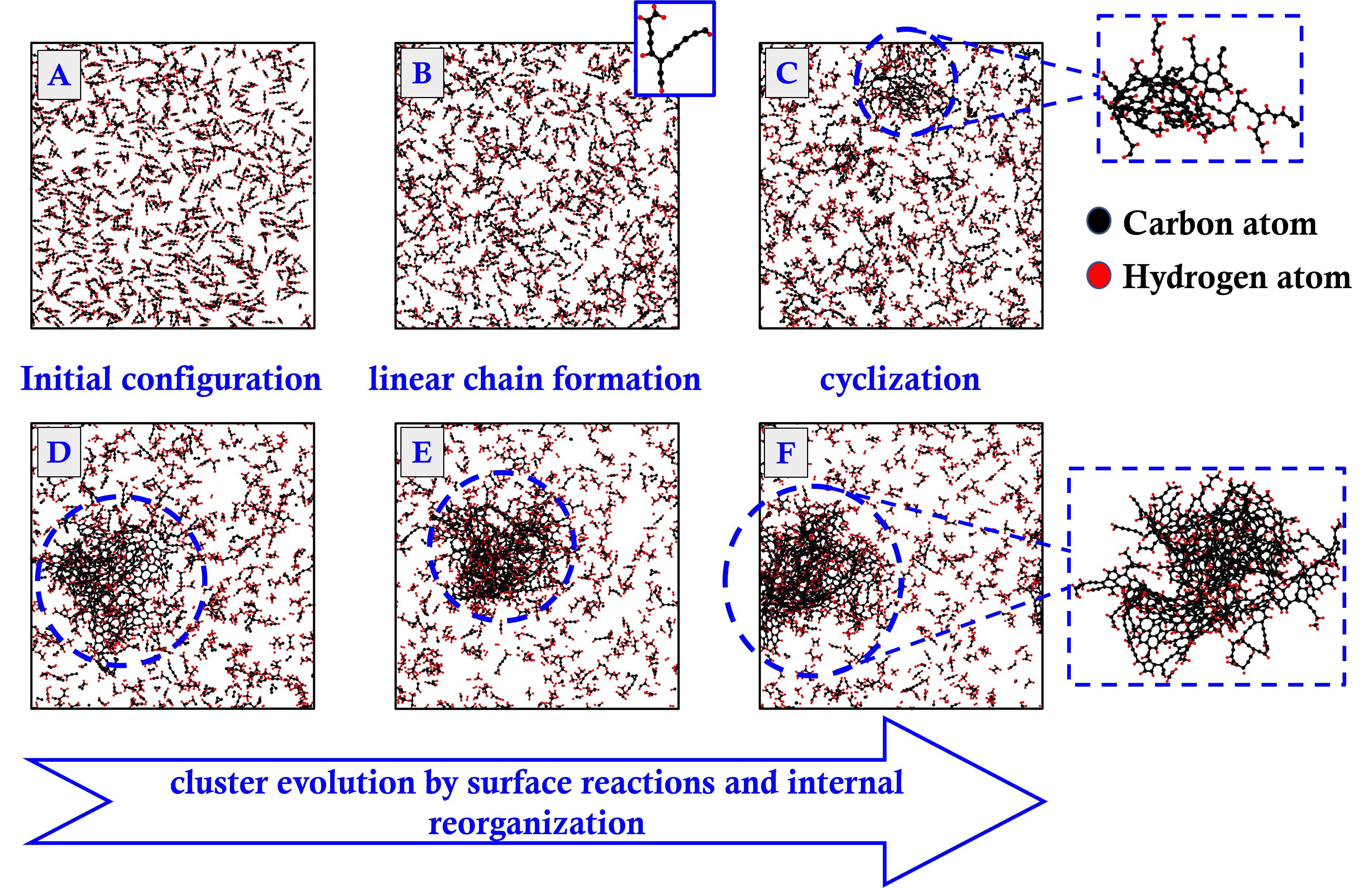}
	\caption{A general representation of steps during the formation and evolution of incipient soot cluster during acetylene pyrolysis (from a simulation performed at 1650 K).}
	\label{f:evolution}
\end{figure}

During the RMD simulations, the system of atoms goes through different
chemical and physical interactions resulting in the formation of larger atomic
clusters due to the pyrolysis of acetylene. The evolution of one of these atomic clusters is depicted in Fig. \ref{f:evolution}. Carbon and hydrogen atoms are represented using black and red dots, respectively. First, the acetylene molecules combine to form small linear chains (Fig. \ref{f:evolution}B: linearization) and then transform into cyclic structures (Fig. \ref{f:evolution}C: cyclization). After cyclization, the small clusters start growing due to both bond formation at the surface and internal reorganization. These larger atomic clusters resemble incipient soot particles (Fig. \ref{f:evolution}D--F). It is important to note that, the collisions are stochastic in nature and the time required for an event, i.e. linearization, cyclization, surface growth. etc., varies based on the initial configurations, and therefore are omitted from the figure for generality. A similar formation mechanism is also reported by Zhang et al. \cite{ Zhang2015Apr} for carbon-black simulations and Sharma et al~\cite{Sharma2021Aug} \editsf{for} acetylene pyrolysis simulations.

\begin{figure}
	\centering
	\includegraphics[width=0.7\linewidth]{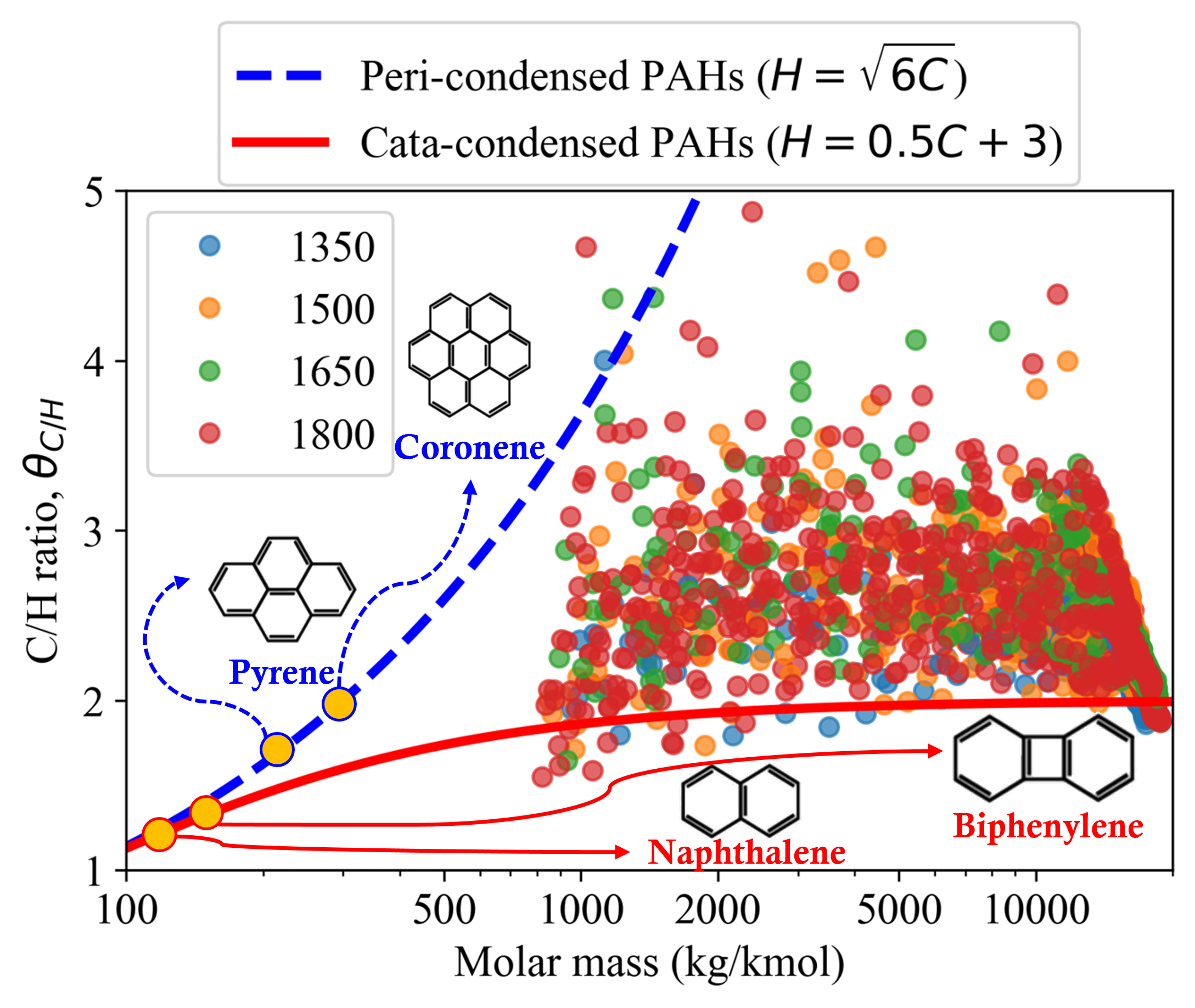}
	\caption{C/H ratio vs. molar mass of soot clusters at different temperatures.}\label{f:pericata}
\end{figure}

The incipient soot clusters are extracted from the RMD simulations at different timesteps to capture the growth. Each simulation is run \editsf{at least five times} with a \editsf{velocity field initialized randomly} in each case to generate soot particles with different evolutionary histories. In total, 3324 individual soot clusters are isolated from the RMD simulations \editsf{(number of carbon atoms ranging from 65 to 1503)}. 
The C/H ratios ($\chr$) of these particles are compared to the theoretical limits for PAHs in Fig.~\ref{f:pericata}.
Based on the compactness, the PAHs can be classified into two categories: (\textit{a}) peri-condensed PAHs, where the carbon atoms in the aromatic structures can be shared by more than two aromatic rings and (\textit{b}) cata-condensed PAHs, where the carbon atoms in the aromatic structure can be shared by at most two aromatic rings. Siegmann and Sattler \cite{Siegmann2000Jan} proposed a relationship between the number of carbon and hydrogen \editsf{atoms} for both peri-condensed and cata-condensed PAHs.
Fig.~\ref{f:pericata} presents the C/H ratio and molar mass of the soot clusters from different temperatures and compares it with the peri-condensed and cata-condensed PAH zones derived from \cite{Siegmann2000Jan}. As observed from Fig. \ref{f:pericata}, the soot clusters fall between the peri-condensed and cata-condensed boundaries, indicating an intricate network of different types of aromatic and aliphatic structures in incipient particles. 
Using atomic force microscopy Commodo et. al. \cite{Commodo2019Jul} showed that smaller aromatic clusters (number of carbon atoms ranging from 6 to 55, lower than what studied in this work) in the early stage of soot formation in a slightly sooting premixed ethylene flame tend to be close the peri-condensed line. \editsf{However, in this work, we find that large clusters lie closer to cata-condensed limit than peri-condensed limit, potentially indicating significant presence of non-aromatic (i.e., aliphatic and alicyclic) structures.}

\subsection{Classification of incipient soot particles} \label{ss:classification}

We tagged, extracted, or calculated the physicochemical features (such as Temperature, number of carbon atoms, number of hydrogen atoms, and molar mass) for each particle. The complete set of features used in this study is listed in~\ref{app_feature}, and two sample particles with the entire feature set are shown in~\ref{app_sample-part}.

\begin{figure}[!htbp]
	\centering
	\subfigure[A t-SNE diagram generated using all the \editsf{3324} incipient soot clusters with  two different k-means clusters.]{%
		\resizebox*{1\linewidth}{!}{\includegraphics{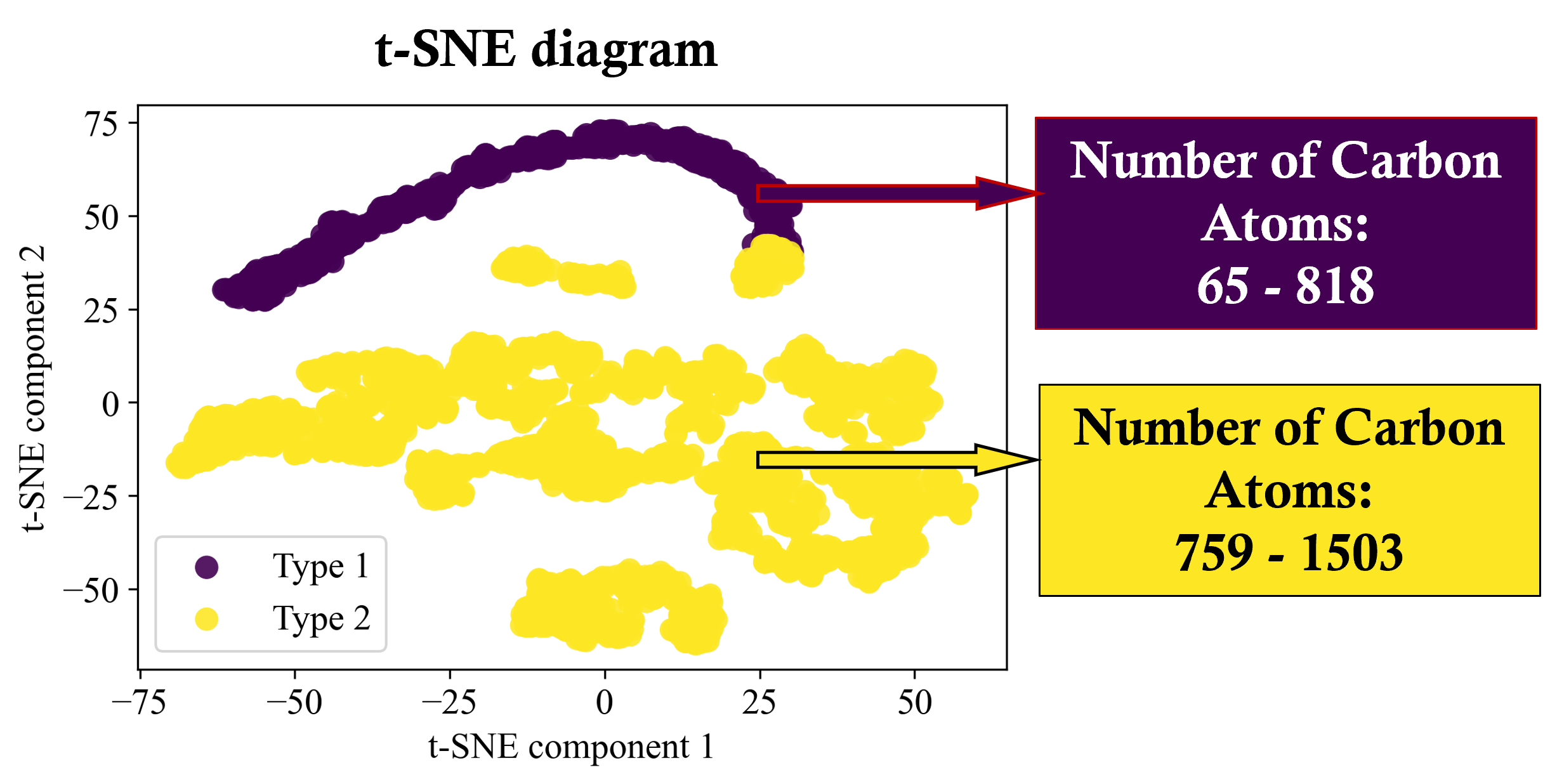}\label{f:kmeans}}}
	\subfigure[Some example particles from each class obtained from RMD simulations. The non-cyclic carbon atom structures are shown in purple dots and the cyclic structures are shown in black. Hydrogen atoms are omitted from the visualization for clarity.]{%
		\resizebox*{0.8\linewidth}{!}{\includegraphics{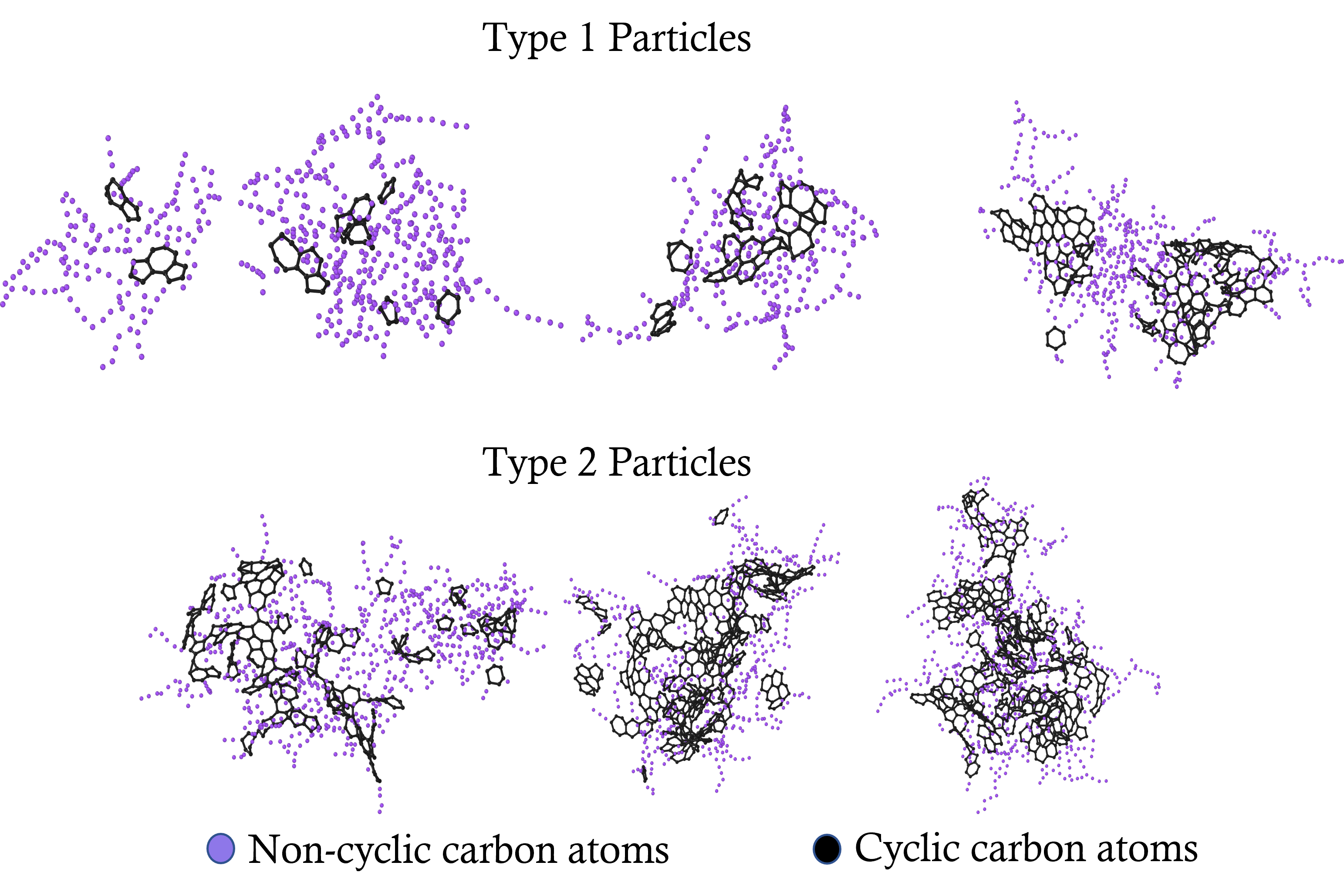}\label{f:sampleParticles}}}
	\caption{Classification of incipient soot particles} \label{f:classification}
\end{figure}

Our initial observation of trends of various internal and physicochemical features revealed a wide variation, indicating that these particles can be classified into multiple groups based on their features. \editsf{The classification of the incipient soot particles is depicted in Fig.~\ref{f:classification}.} We attempted two unsupervised machine-learning techniques to extract unique classifications that may exist in the incipient particle sample space. The first method is the k-means clustering algorithm \cite{Lloyd1982Mar}, which is used to label particles of different classes based on all extracted features of the particles. Then we used the t-dispersion stochastic neighbor embedding (t-SNE)\cite{vanderMaaten2008} plot to display the particle properties on a 2D map. In the t-SNE diagram, similar clusters (i.e., potentially belonging to the same class) are expected to be close to each other. Although the number of classes was not known a priori, trial and error with k-means clustering revealed good results with two classes. For identification purposes, these two classes are referred to as ``\texttt{type~1}'' and ``\texttt{type~2}'' particles, respectively. The resulting t-SNE diagram is shown in Figure~\ref{f:kmeans}. 

Looking closely at the two classes, we see that the particles exhibiting similarity fall into a nearly continuous size range. For example, in the first class, (\texttt{type~1}) the incipient particles have a lower total number of carbon atoms ($65-818$) whereas in the second class (\texttt{type~2}) the particle have a higher number of total atoms ($759-1503$). This essentially points to the fact that the characteristics of the incipient particles change after a certain level of growth: smaller particles (type~1) show different features and trends than larger particles (type~2). It should be noted here that the number of total carbon atoms is not a unique marker of the threshold between type~1 and type~2 -- as indicated by a small overlap in the number of carbon atoms range between the types -- but acts as a very good surrogate for threshold identifier. In total, we have obtained 670 type~1 and 2654 type~2 incipient particles from a total of 3324 particles. Fig. \ref{f:sampleParticles} depicts some example particles from the analyzed sample space. Here, the non-cyclic carbon atom structures are shown in \editsf{purple} dots and the cyclic structures are shown in black, \editsf{while} hydrogen atoms are omitted from the visualization for clarity.

\subsection{Comparison with experimental data} \label{ss:validation} 
The particles obtained in the current study show very good match with experimentally observed properties of incipient soot particles as reported in the literature. For example, the mean density of the particles is our study is calculated to be $1.53 \pm 0.08$~g/cm$^3$, which is is an excellent match with the empirical soot density of by Johansson et al~\cite{Johansson2017Dec}, who reported the value to be 1.51~g/cm$^3$. More detailed comparison of average particle properties are presented in~\cite{Mukut_paper2}. 

\subsection{Internal structure of incipient soot particles} \label{ss:internal_structure}

\begin{figure}[!htb]
	\centering
	\subfigure[]{%
		\resizebox*{0.49\linewidth}{!}{\includegraphics{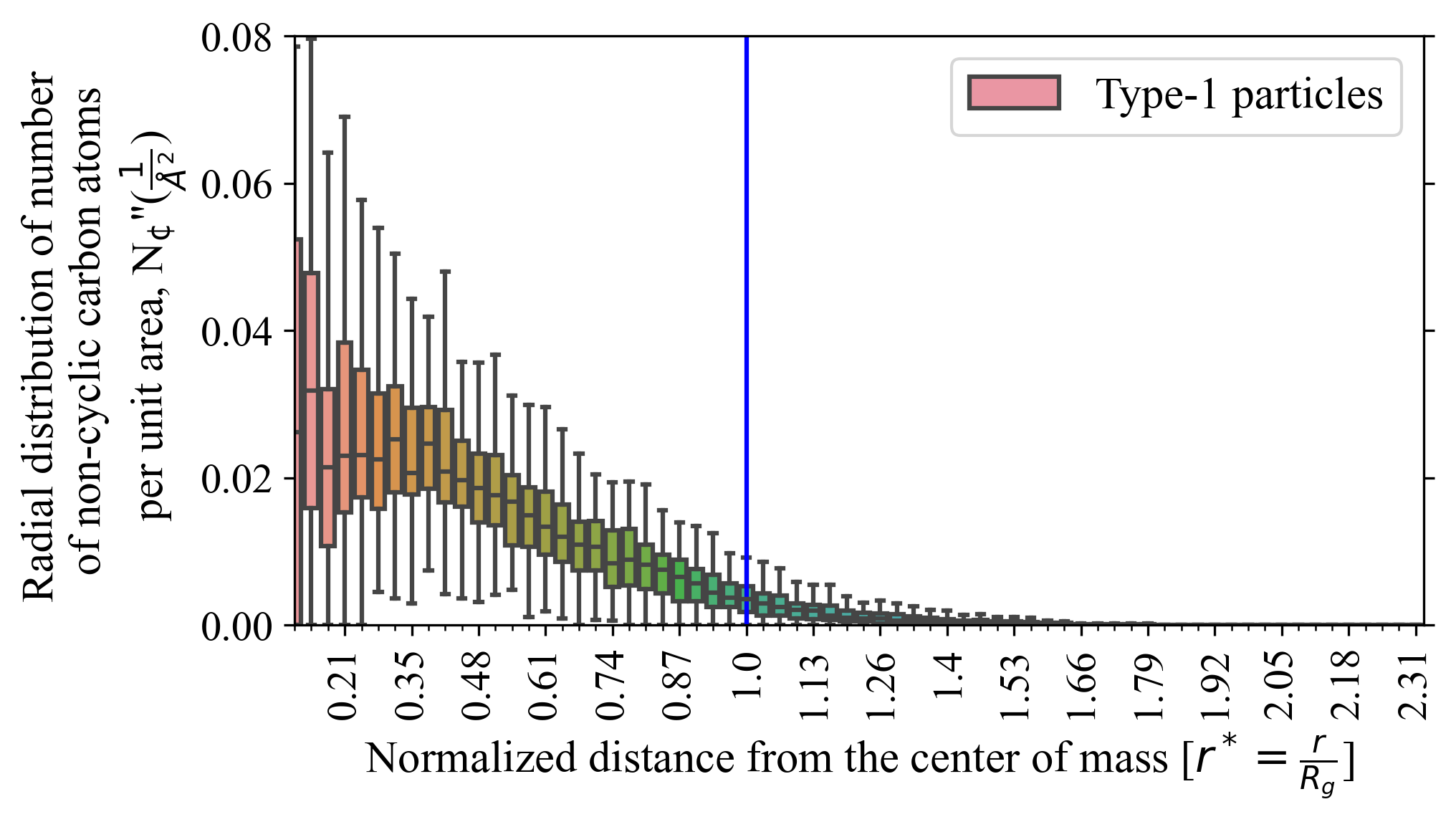}\label{f:sali1}}}
	\subfigure[]{%
		\resizebox*{0.49\linewidth}{!}{\includegraphics{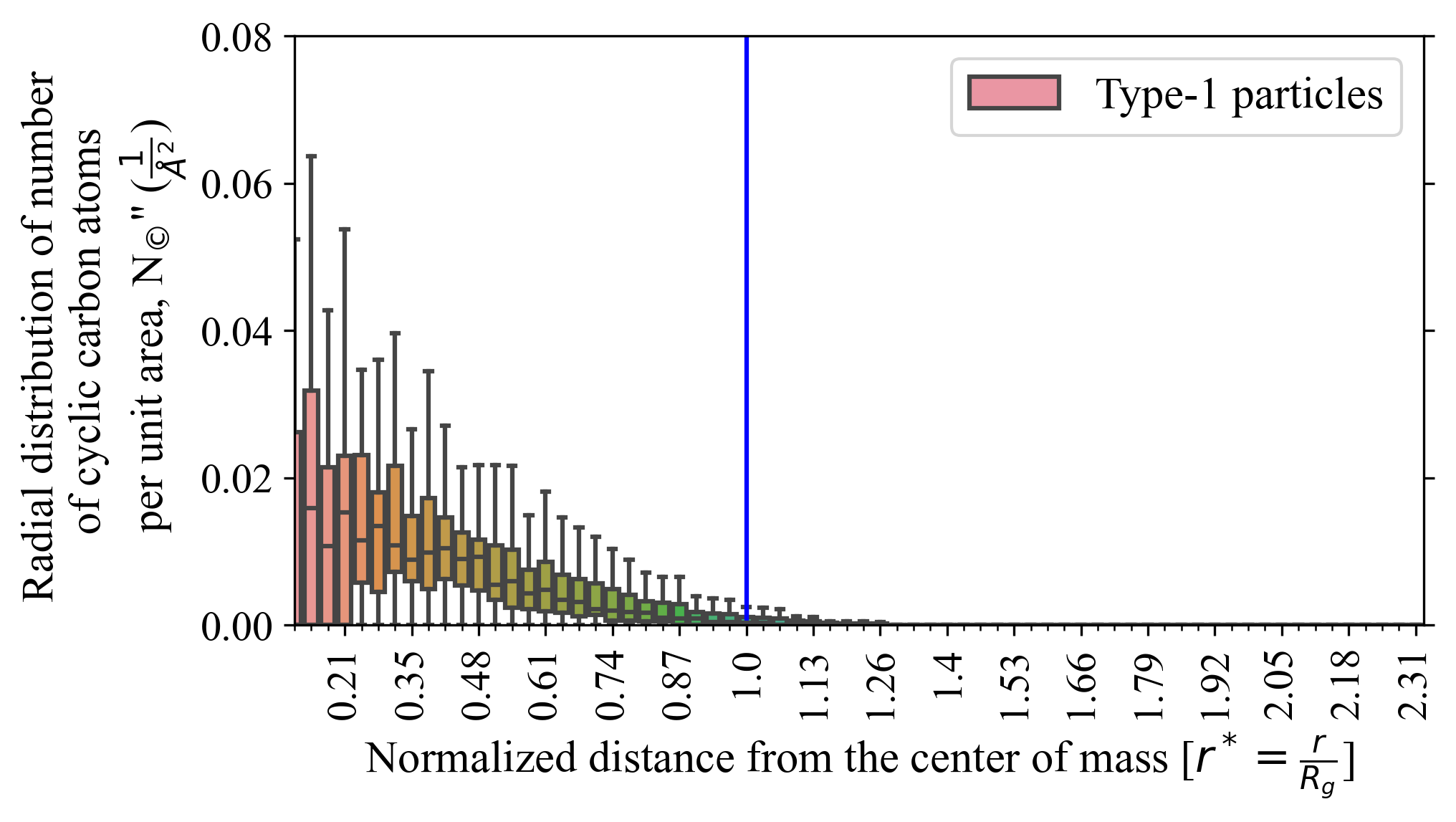}\label{f:saro1}}}
	\subfigure[]{%
		\resizebox*{0.49\linewidth}{!}{\includegraphics{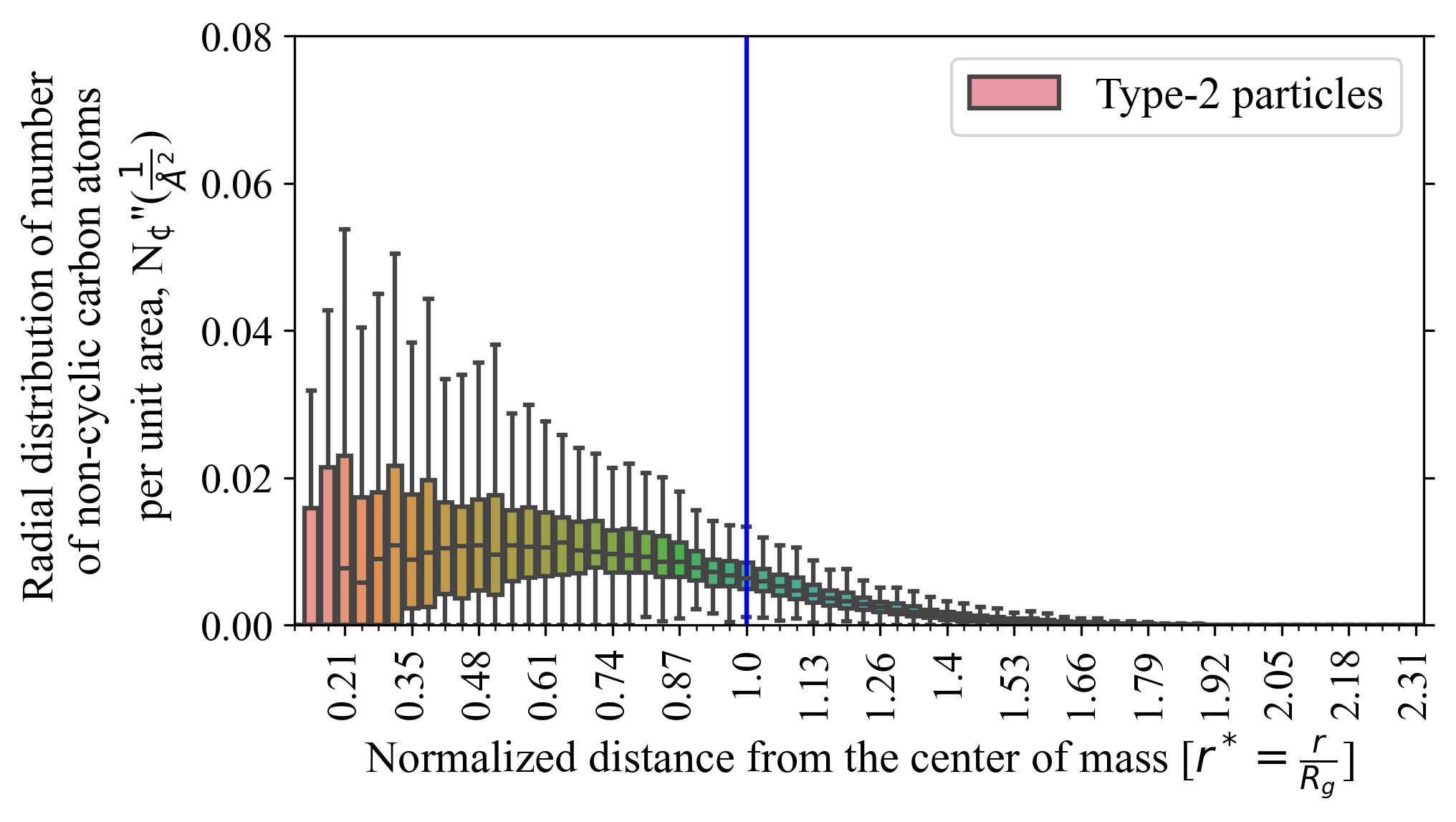}\label{f:sali2}}}
	\subfigure[]{%
		\resizebox*{0.49\linewidth}{!}{\includegraphics{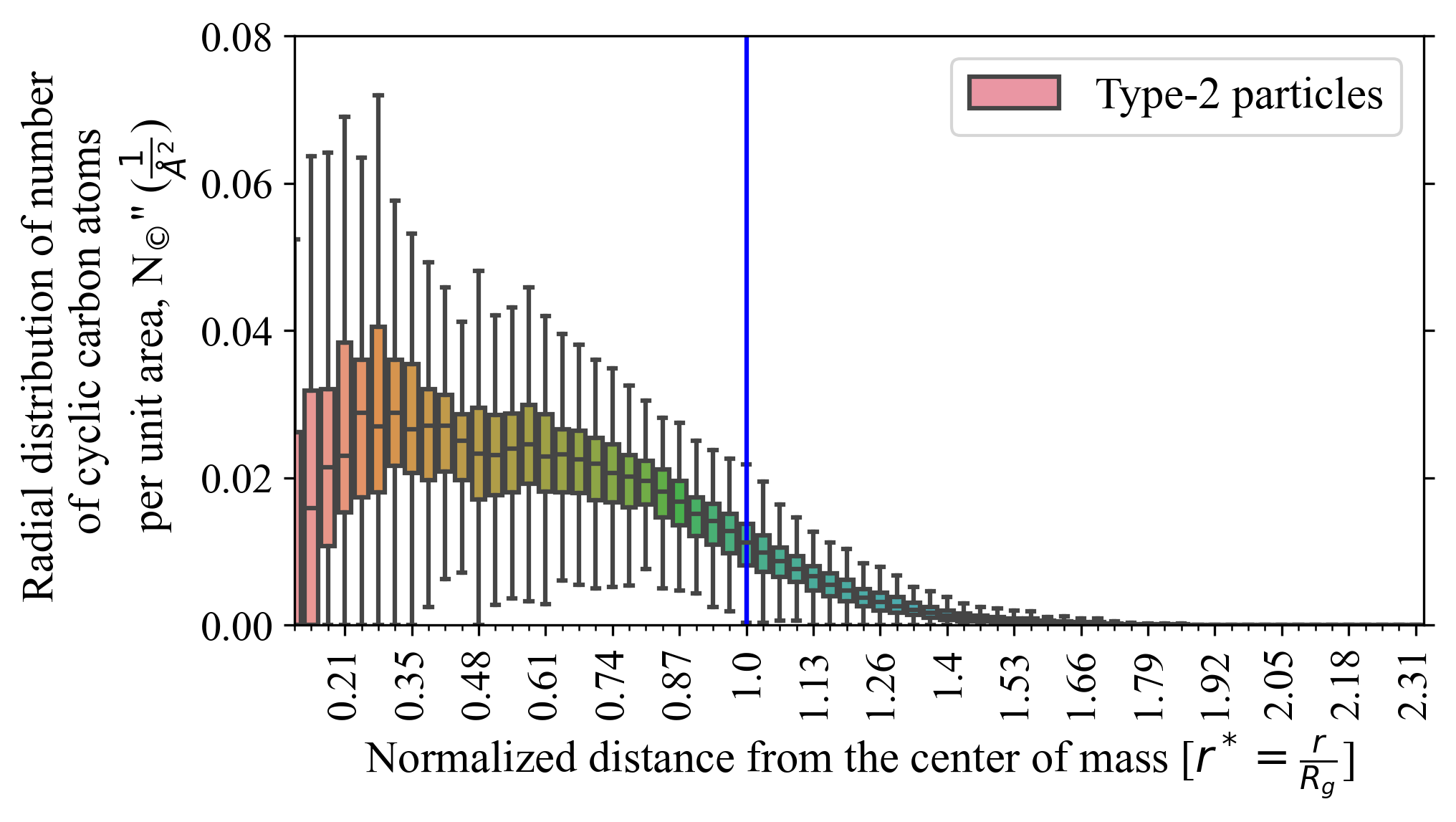}\label{f:saro2}}}
	\caption{Radial distribution of (a,c) non-cyclic and (b,d) cyclic carbon atoms ($\NcnonringP(r^*)$ and $\NcringP(r^*)$)  in (a,b) type 1 and (c,d) type 2 soot particles as a function of the normalized radial distance ($r^*=\frac{r}{R_g}$) from the center of mass. The blue vertical line is at a radial distance equivalent to $R_g$. \label{f:surfDensity}}
\end{figure}

The radial distributions of various quantities relevant to the analysis of the internal structure of the incipient particles are presented as box and whiskers plots in subsequent figures. In this visualization, every box includes the data points within the second and third quartiles, the horizontal line inside the box indicates the median value and whiskers represent the range of the data. 
The statistics were found to be insensitive to the process temperature. Hence  aggregate data for all temperatures is presented here.

The radial distribution of cyclic and non-cyclic carbon atoms per unit area for type~1 (Figs.~\ref{f:sali1}-\ref{f:saro1}) and type~2 (Figs. \ref{f:sali2}-\ref{f:saro2})
incipient particles as a function of the normalized radial distance from the
center of mass is shown in Fig. \ref{f:surfDensity}. The blue vertical line
depicts the location where the radial distance becomes equal to the radius of
gyration ($R_g$) of individual particles.
For type~1 particles, an abundance of non-cyclic carbon atoms is observed in
the \editsf{central (i.e., less than 50\% of $R_g$)} region. The number of non-cyclic carbon
atoms is almost twice the number of carbon atoms in cyclic structures in
type~1 particles. Almost all the carbon atoms reside near the
central region of type~1 particles and the number of carbon atoms
quickly drops to zero as we go away from the center of mass. This indicates
that in type~1 particles, i.e., at the very early stages of soot formation, the number
of non-cyclic structures is significantly higher than the number of cyclic
structures.

The type~2 particles, on the other hand, show a different trend where a very
concentrated region of cyclic carbon atoms is observed in the central region
of the particles. The concentration of cyclic carbon atoms slowly decreases as the distance from the center of mass increases.
The number of non-cyclic carbon atoms increases from a very low value in the
central region, then reaches a steady value \editsf{near} the radius of gyration
and then gradually drops to zero as the distance increases beyond the radius of gyration. This indicates that the
non-cyclic carbon atoms are more likely to be present
in the outer region (what can be presumably considered near the particle surface) of the type~2 incipient
particles while the central region is dominated by the cyclic carbon atoms.

\begin{figure}[!htb]
	\centering
\includegraphics[width=0.49\linewidth]{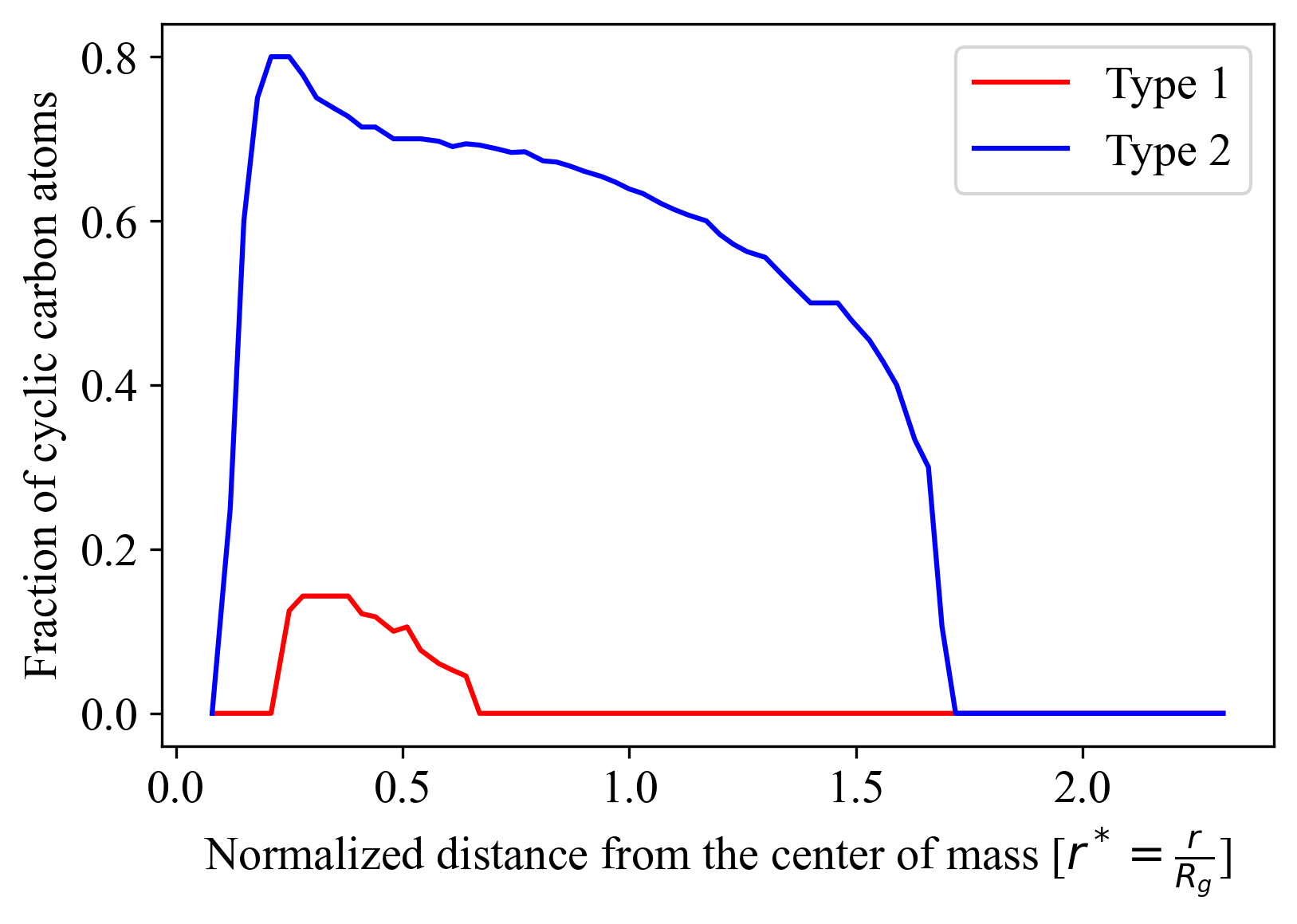}
	\caption{Radial distribution of the median fraction of cyclic carbon.} \label{f:cycl_fr}
\end{figure}
\editsf{If the radial distribution of the median value of the fraction of cyclic carbon atoms with respect to total carbon atoms is analyzed (shown in Fig.~\ref{f:cycl_fr}), the dominance of cyclic carbon atoms in type 2 particles becomes very clear. Figure~\ref{f:cycl_fr} shows a prominent central region where more than half of the carbon atoms belong to a ring structure. These evidences} suggest that the internal chemical structure of
incipient soot particles changes as the particles transition from type~1 to
type~2. This indicates the development of core-shell structures as the incipient soot grows and matures. The presence of such core-shell structure has been theorized in the literature. For example, Michelsen et. al. \cite{Michelsen2022Aug} used a fractal core-shell model to explain the changes in the structure of soot aggregates and primary particles at different heights of a laminar co-flow ethylene-air flame. 
Kholghy et. al. \cite{Kholghy2016Apr} proposed a surface shell formation model to predict maturity of soot primary particles. More directly, recently Botero et. al. \cite{Botero2019Jan} studied the internal structure of soot particles using high-resolution transmission electron microscopy (HRTEM) to identify the PAH structures in the core and shell regions, and suggested the presence of a stabilized core region indicating nano-structural mobility. 
Pascazio et. al. \cite{Pascazio2020Sep} utilized RMD simulation and identified different levels of crosslinking in core and shell in hypothetical soot particles. \editsf{Kelesidis et al. \cite{Kelesidis2023Dec} also investigated oxidation dynamics of carbonaceous nanoparticles having various core-shell structures using lattice Monte Carlo simulations.} The clear difference in the radial distribution of cyclic and non-cyclic carbon atoms between type 1 and type 2 particles supports these findings.

\begin{figure}[!htbp]
	\centering
	\subfigure[]{%
		\resizebox*{0.7\linewidth}{!}{\includegraphics{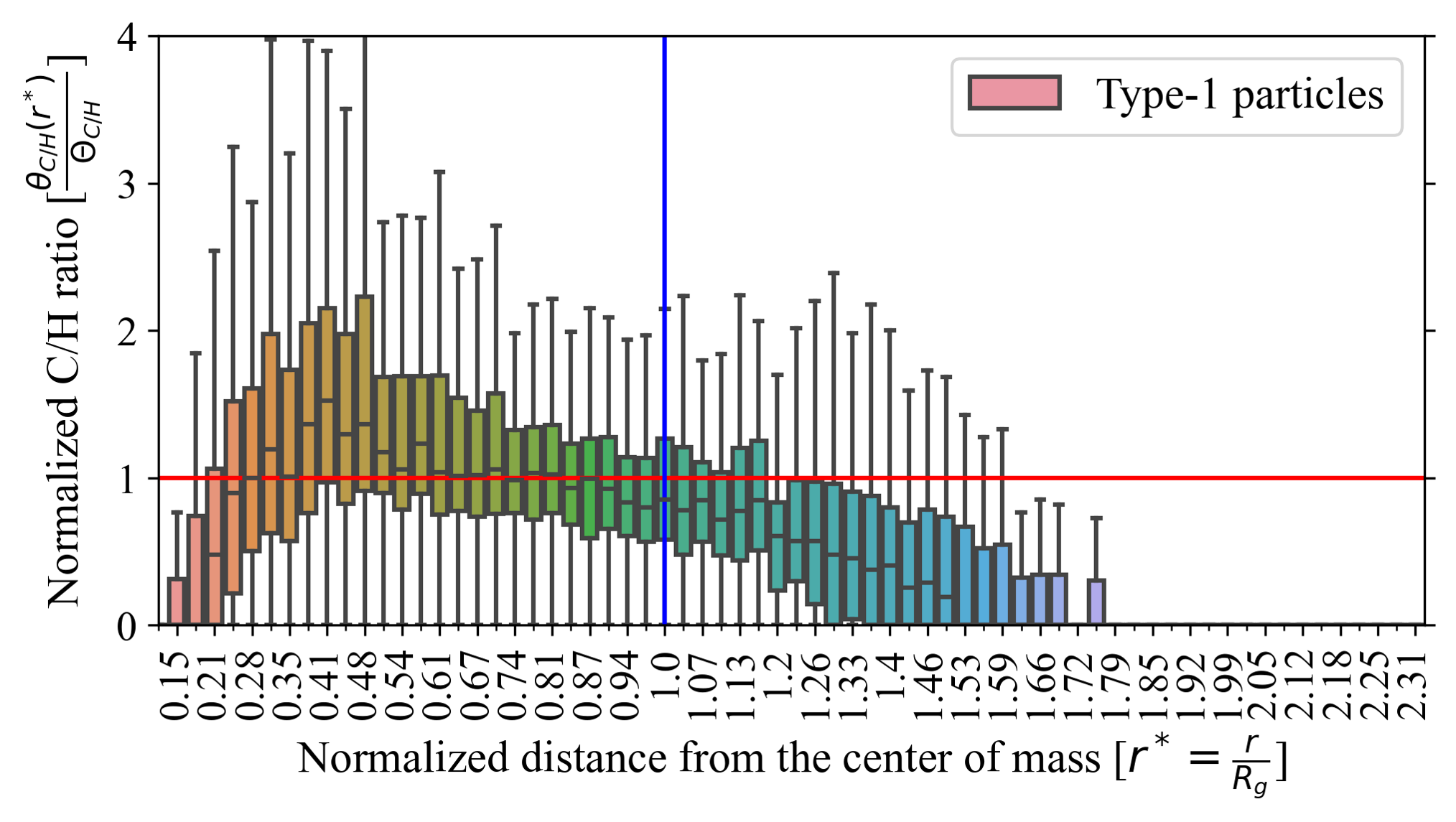}\label{f:ch1}}}
	\subfigure[]{%
		\resizebox*{0.7\linewidth}{!}{\includegraphics{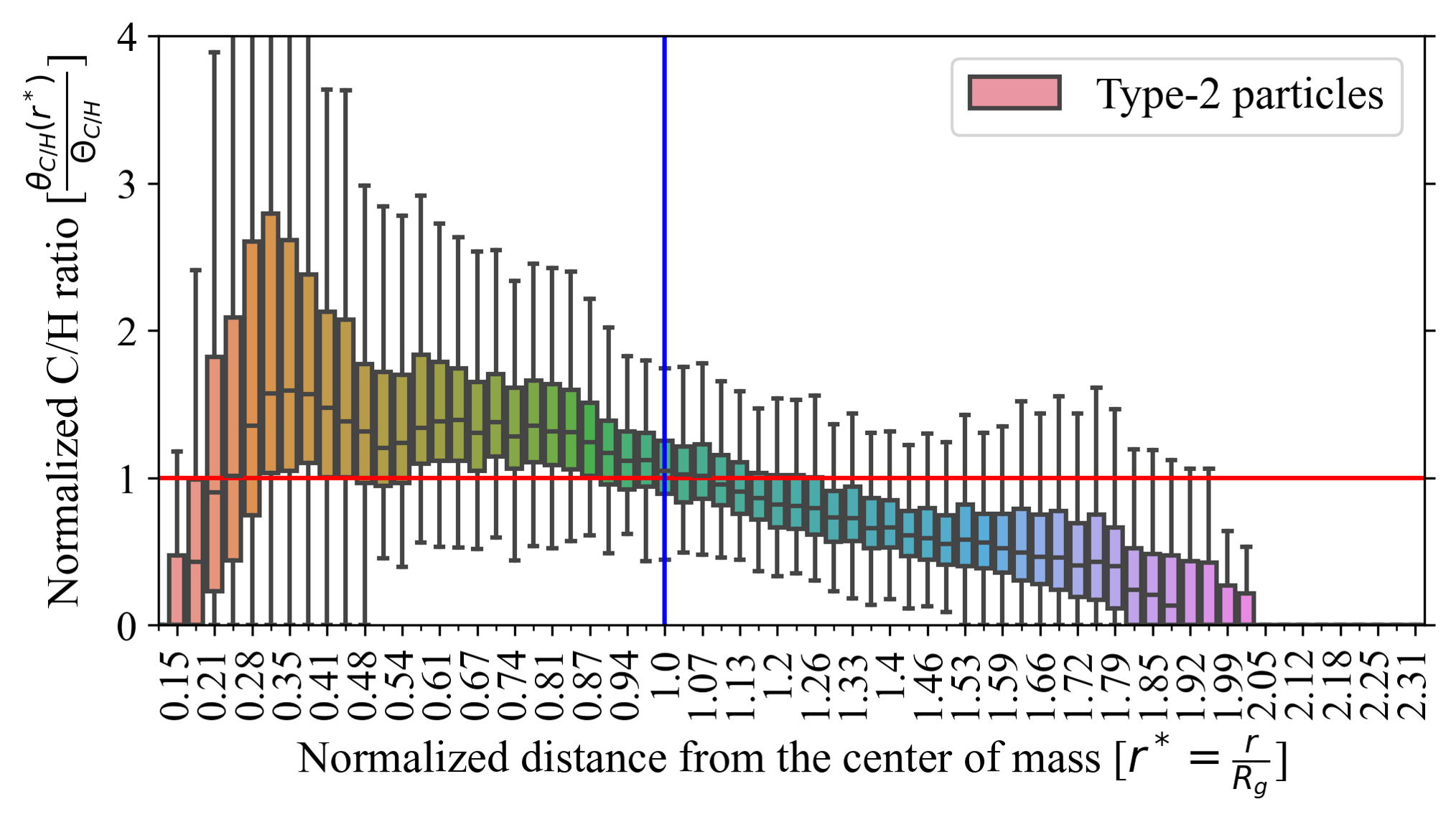}\label{f:ch2}}}
	\caption{Radial distribution of normalized C/H ratio ($\chrnorm(r^*)= \frac{\chrl(r^*)}{\chr}$) in incipient particles as a function of normalized radial distance ($r^*=\frac{r}{R_g}$)  from the center of mass. The blue vertical line is at a radial distance equivalent to $R_g$. The $\frac{C}{H}$ ratio of the strip and the particle are the same along the red horizontal line.} \label{f:CHDist}
\end{figure}

The changes in chemical properties inside the incipient soot particles can also
be observed in the radial distribution of carbon to hydrogen ratio. 
The normalized local C/H ratio for type~1 (Fig. \ref{f:ch1}) and
type~2 (Fig. \ref{f:ch2}) particles are plotted as a function of the normalized radial distance from the center of mass in Fig. \ref{f:CHDist} using box and whisker plots. The blue vertical line depicts the location where the radial distance becomes equal to the radius of gyration ($R_g$) of individual particles. The red horizontal line indicates where the local C/H ratio is equal to the particle C/H ratio.

The normalized local C/H ratio increases up to a
certain distance from the center of mass and then starts to drop. For type~1
particles, the increase in the local C/H ratio \editsf{take longer distance from the center of mass}, and the median value
gradually reaches a peak value slightly higher than the particle C/H ratio
(about 1.3 times). After that, the local C/H ratio starts to drop and the median value reaches a
value close to the particle C/H ratio at around 65\% of the radius of gyration.
The value stays close to the particle C/H ratio up to the radius of gyration,
and then slowly drops to zero. This indicates a very small or \editsf{no} dense core region in type~1 particles. The
demarcation between the core and shell regions is not clear in
type~1 particles because of the absence of a pronounced core \editsf{as the local C/H ratio remains close to the particle C/H ratio}. For type~2 particles, however, the local C/H ratio increases
rapidly in the central region and the median value reaches a peak value of
about 1.7 times the particle C/H ratio. After that, the local C/H ratio
starts to drop and reaches a value equal to the global C/H ratio at the radius
of gyration. Unlike type~1 particles, the region where the median of the
normalized local C/H ratio is close to unity is very narrow in type~2
particles. This
indicates that the dense core region of type~2 particles is well-developed and
the \editsf{separation} between the core and shell regions is more pronounced than in
type~1 particles.

\begin{figure}[!htb]
	\centering
	\subfigure[]{%
		\resizebox*{0.7\linewidth}{!}{\includegraphics{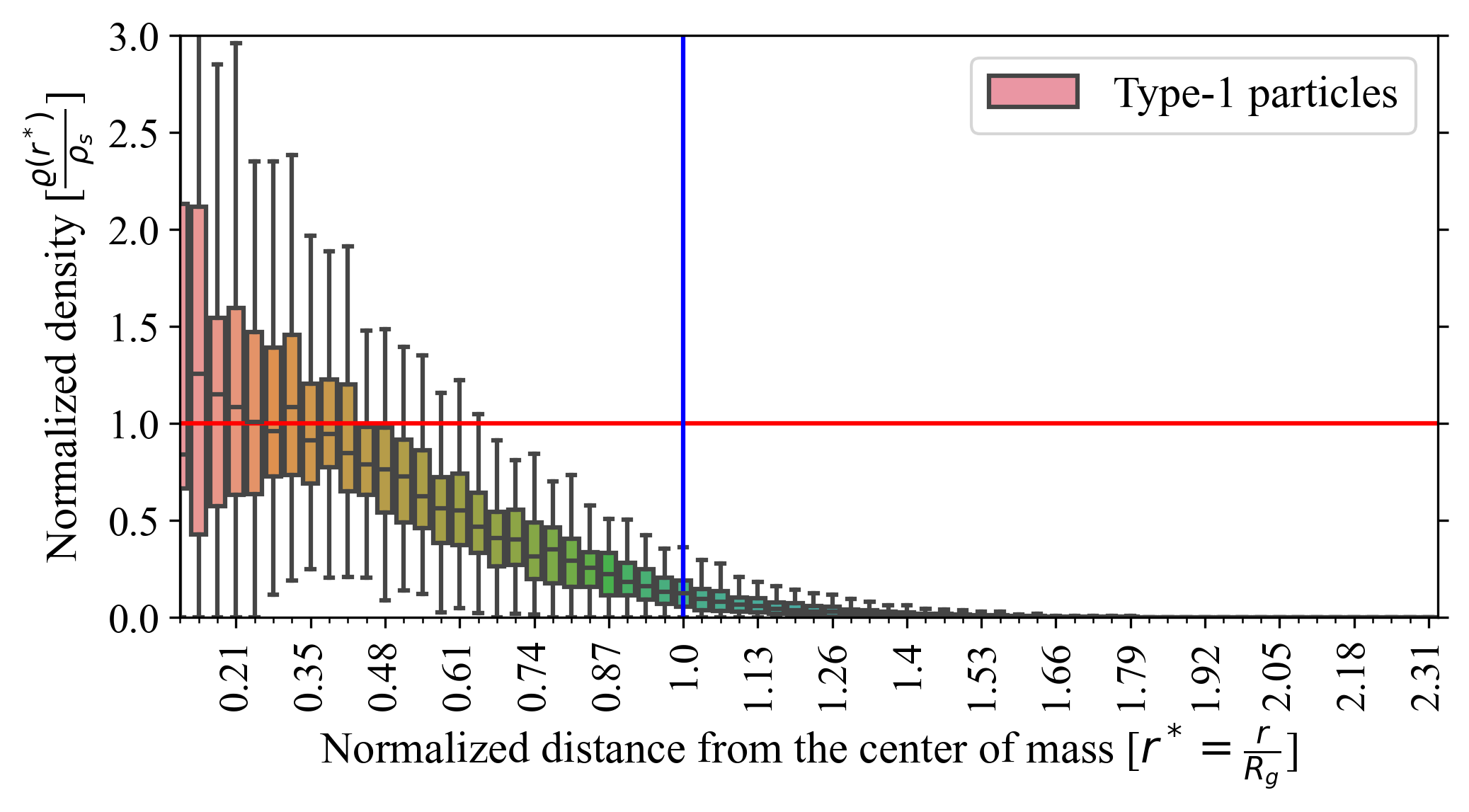}\label{f:d1}}}
	\subfigure[]{%
		\resizebox*{0.7\linewidth}{!}{\includegraphics{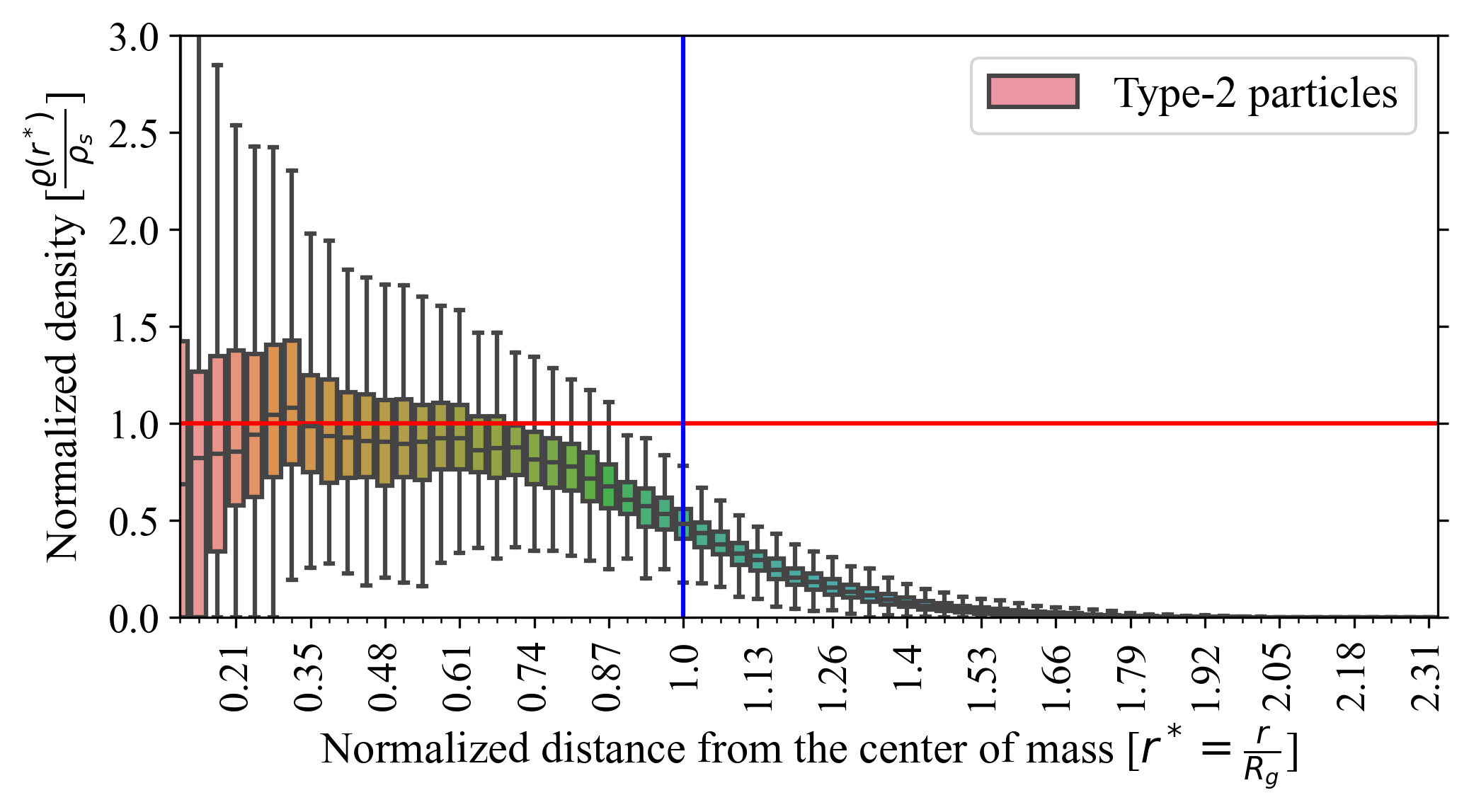}\label{f:d2}}}
	\caption{Radial distribution of normalized local density ($\rho^*(r^*) = \frac{\varrho(r^*)}{\rho_s}$) inside incipient particles as a function of normalized radial distance ($r^*=\frac{r}{R_g}$)  from the center of mass. The blue vertical line is at a radial distance equivalent to $R_g$. The density  of the strip and the particle are the same along the red horizontal line.} \label{f:densityDist}
\end{figure}

The radial distribution of the normalized density inside type~1 (Fig.
\ref{f:d1}) and type~2 (Fig. \ref{f:d2}) incipient soot particles is presented
in Fig. \ref{f:densityDist} as a function of the normalized radial distance from the center of mass of soot particles ($r^*$) as box and whisker plots. 
For type~1 particles, Fig. \ref{f:densityDist} shows a very small dense central
region that extends up to about 40\% of the radius of gyration. The local
density of type~1 particles is maximum near the center of mass, and it drops
gradually as the distance from the center of mass increases. For type~2
particles, on the other hand, the dense core region is larger and extends up to
about 50-60\% of the radius of gyration. The local density of type~2 particles
stays close to the particle density up to around 60\% of the radius of gyration,
and then it quickly drops.

\subsection{The boundary between the core and shell} \label{ss:proposition}
\begin{figure}[!htb]
	\centering
	\subfigure[]{%
		\resizebox*{0.48\linewidth}{!}{\includegraphics{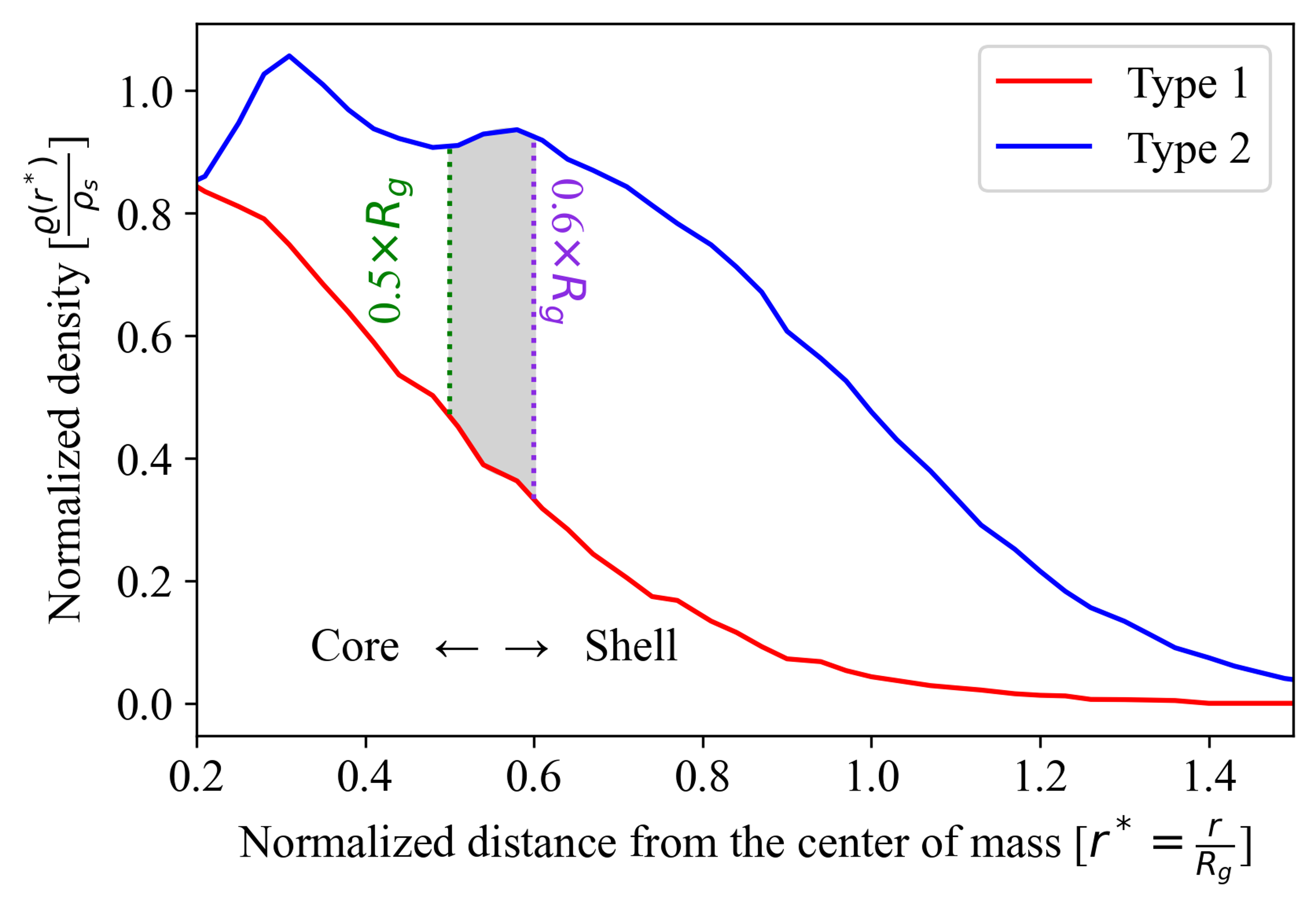}\label{f:median_density}}}
  \hfill
	\subfigure[]{%
		\resizebox*{0.48\linewidth}{!}{\includegraphics{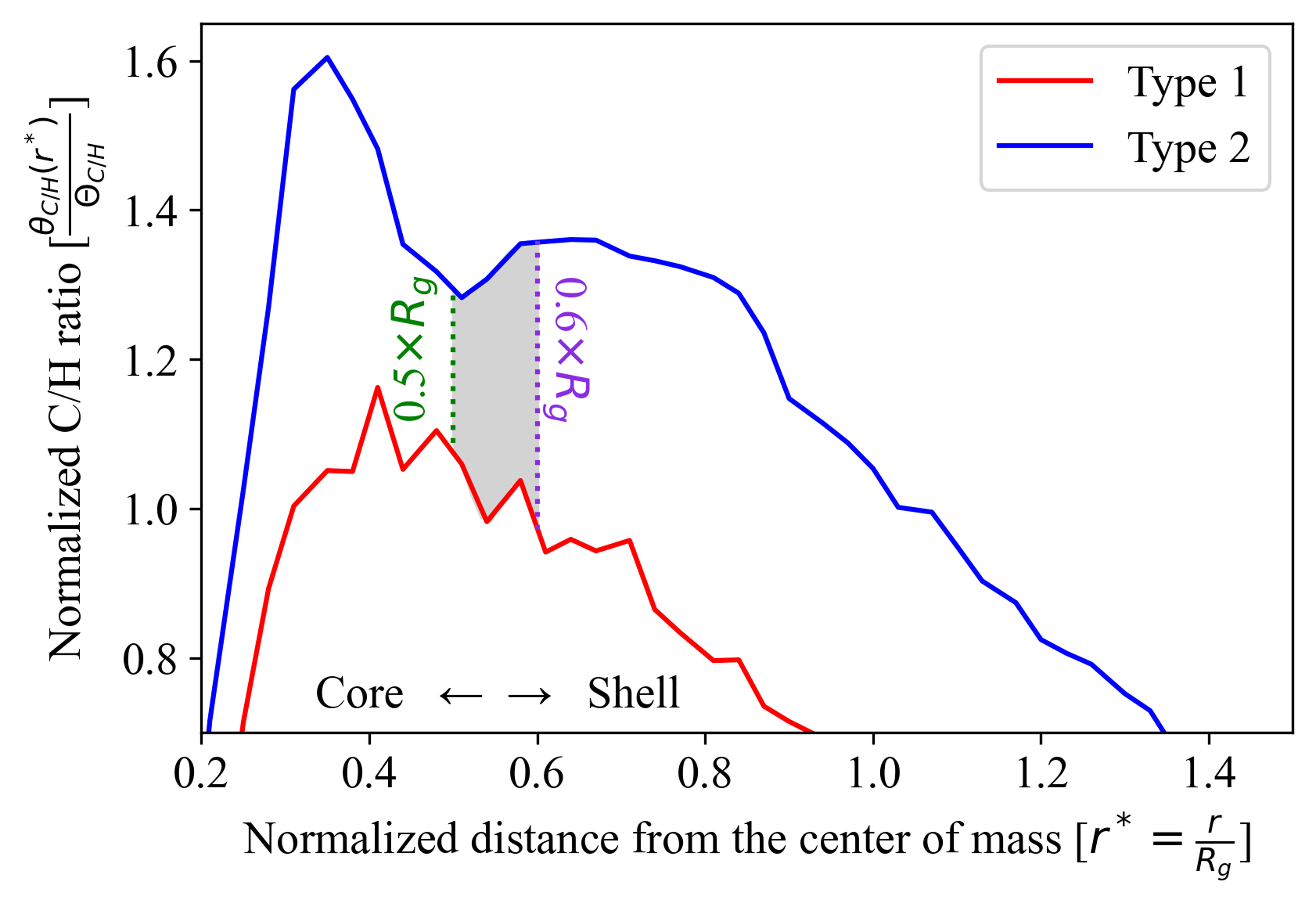}\label{f:median_ch}}}
	\subfigure[]{%
		\resizebox*{0.48\linewidth}{!}{\includegraphics{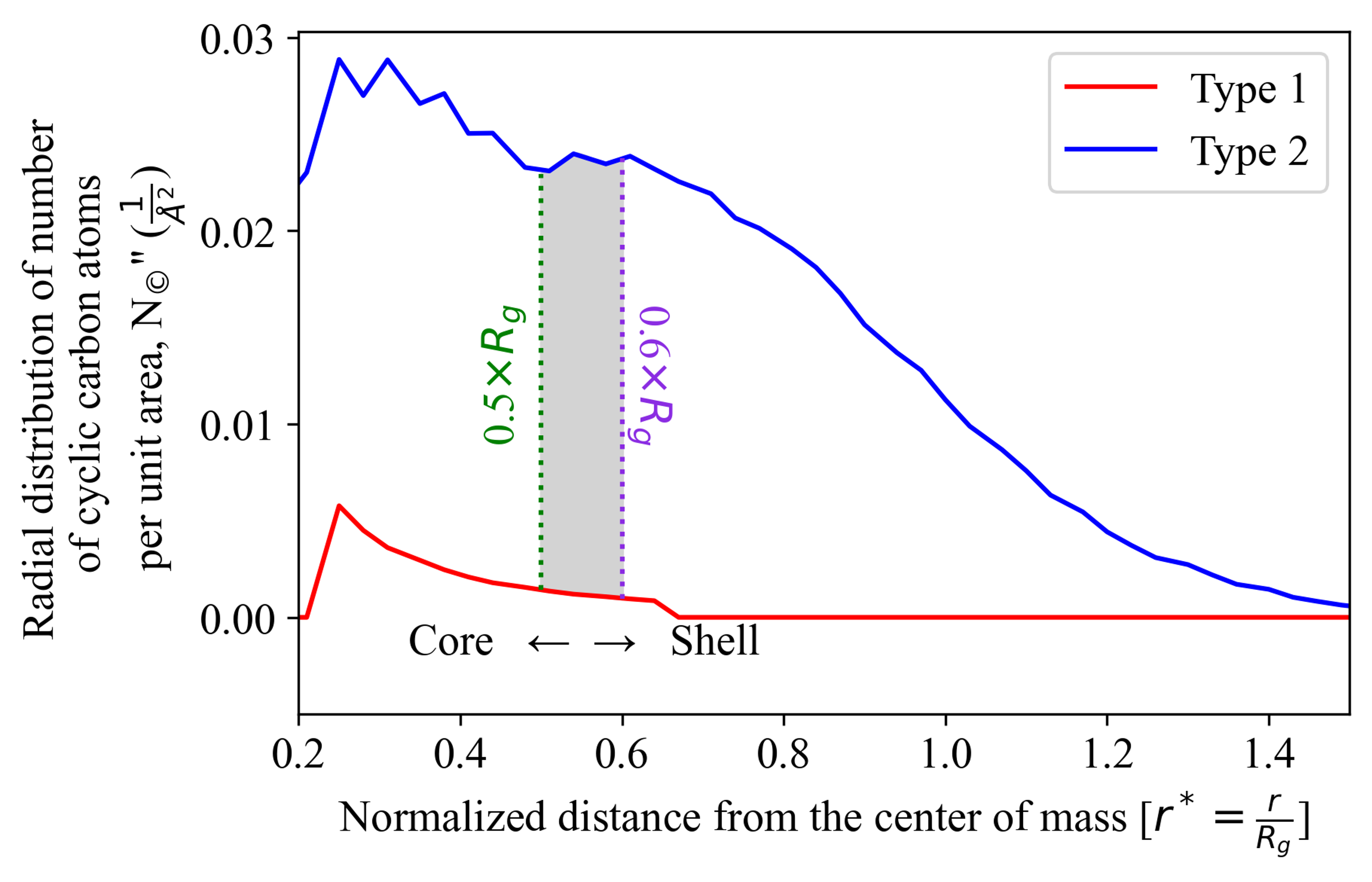}\label{f:median_aro}}}
\hfill
 \subfigure[]{%
		\resizebox*{0.48\linewidth}{!}{\includegraphics{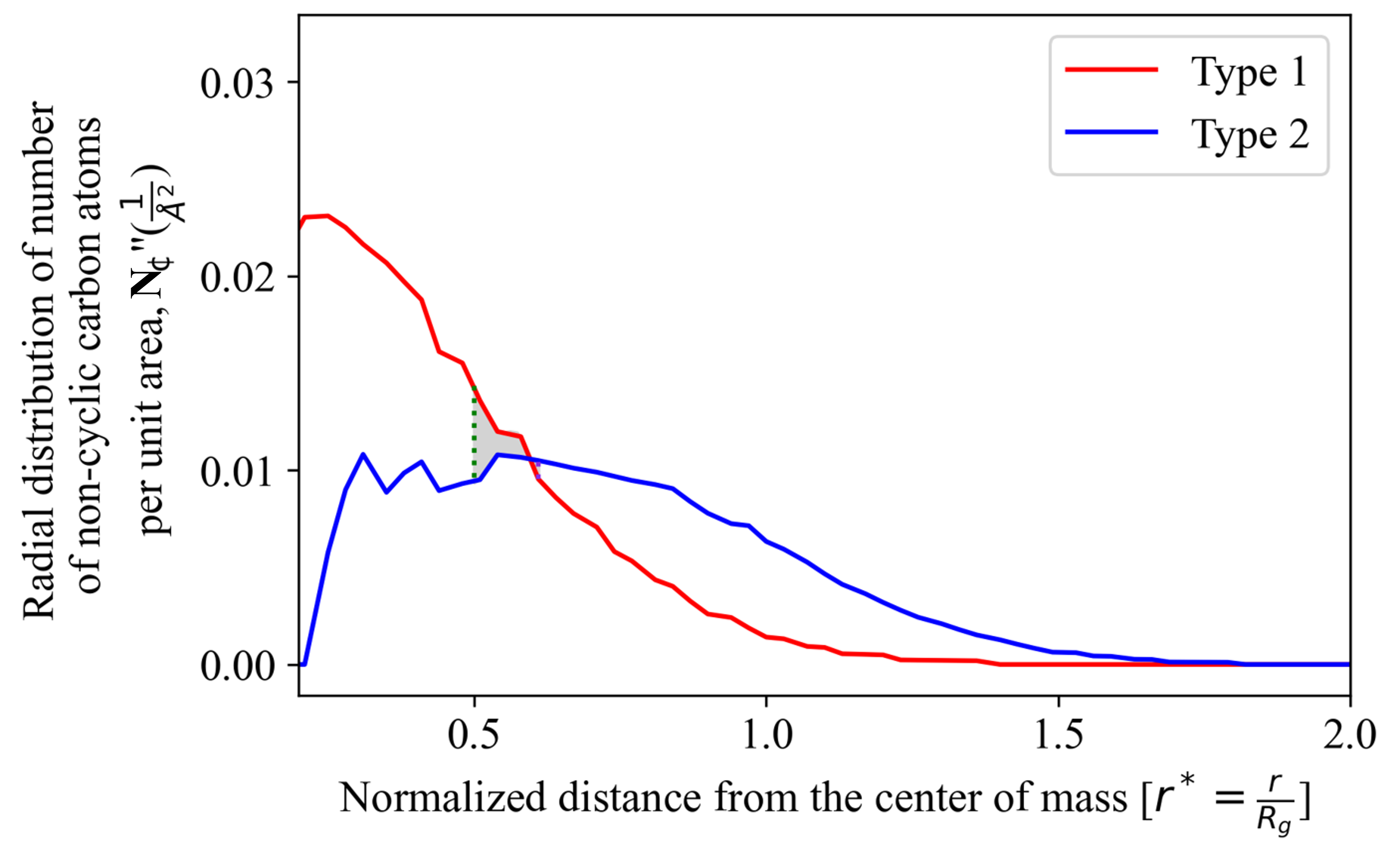}\label{f:median_ali}}}
	\caption{Identification of core and shell based on the radial distribution of normalized medians (a) density, (b) C/H ratio and radial distribution of the number of (c) cyclic and (d) non-cyclic carbon atoms per unit area in incipient particles as a function of normalized radial distance ($r^*=\frac{r}{R_g}$) from the center of mass. \label{f:median_properties}}
\end{figure}

The results discussed so far indicate the presence of a core and shell structure in incipient soot particles. This is further examined by looking at the median values of normalized local density ($\rho_s^*$), normalized C/H ratio ($\chrnorm$), and radial distribution of carbon atoms ($\NcringP$ and $\NcnonringP$) in Fig.~\ref{f:median_properties} for both type~1 (in red) and type~2 (in blue) incipient soot particles. No significant trend is noticed from the radial distribution of non-cyclic carbon per unit area (Fig. \ref{f:median_ali}). However, the normalized density (Fig. \ref{f:median_density}), C/H ratio (Fig. \ref{f:median_ch}) and the radial distribution of cyclic carbon atoms per unit area (Fig. \ref{f:median_aro}) demonstrate unique common trends in type~2 particles, which is not observed in type~1 particles. All of the quantities
\begin{enumerate}
	\item reach a maximum value near the center of mass and decrease gradually to a local minimum at a distance of about 50\% of the radius of gyration,
	\item show the presence of a plateau region between 50\% and 60\% of the radius of gyration, 
	\item and drop monotonously after approximately 60\% of the radius of gyration. 
\end{enumerate}

\begin{figure}[!htbp]
	\centering
	\includegraphics[width=0.65\textwidth]{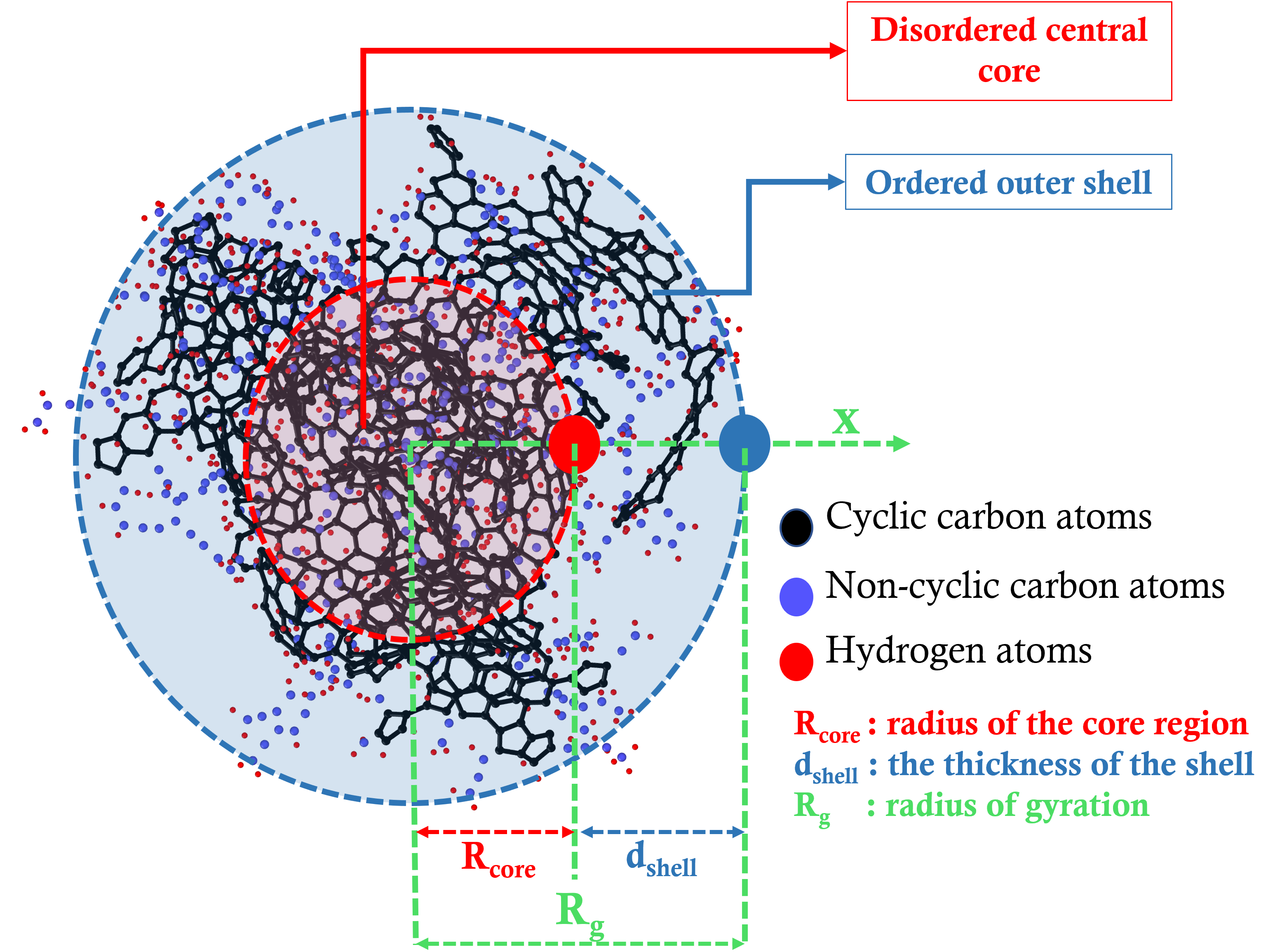}
	\caption{Schematic of core and shell structure of incipient particles.}\label{f:core-shell-boundary}
\end{figure}

The first region or the central region can be identified as the \textit{core} of the soot as it is near the center of mass, \editsf{it is} denser and contains more rings than other regions \editsf{(as seen in Fig. \ref{f:saro1} and \ref{f:saro2})}.
The local minima mark the beginning of the boundary between the \textit{core} and the \textit{shell}. \editsf{The narrow plateau region can be thought of as the boundary region between \textit{core} and \textit{shell} regions}. And finally, the gradual descent of these quantities indicates the \textit{shell} region. In the type~1 particles, we can only see the gradual descent stage, indicating that the core-shell demarcation is not yet developed, i.e., there is no developed core.  The reason for such a trend in density, C/H ratio and ring structures can be explained by the nature of the stacking of cyclic molecules (disordered and ordered), as shown in Fig. \ref{f:core-shell-boundary}. As observed in the schematic in Fig. \ref{f:core-shell-boundary}, the core region comprises an interconnected cross-linked network of cyclic molecules while the shell region contains sheet-like organization of cyclic molecules. Such structural differences in core and shell are also supported by results presented by Pascazio et al. \cite{Pascazio2020Sep}. \editsf{The core size of $0.5-0.60R_g$ obtained here is also consistent with the core size of $0.5-0.75R_g$ suggested by lattice Monte Carlo simulations~\cite{Kelesidis2023Dec}.}

\section{Conclusion} \label{conclusion}

A series of reactive molecular dynamics (RMD) simulations were performed to study the evolution of incipient soot particles during acetylene pyrolysis at four different temperatures.  A total of 3324 incipient soot particles were obtained at different stages of evolution from these simulations. The mass, volume, surface area, radius of gyration, density, C/H ratio, and the number of \editsf{cyclic structures} were calculated for each particle. 
The internal structures of RMD-derived soot and their classification were investigated. Using unsupervised machine-learning techniques, the incipient soot particles are classified into two types -- type~1 and type~2 -- based on their morphological and chemical features. This classification was found to be very well predicted by the size of the particles with smaller particles in the type~1 class and larger particles in type~2.
The internal structures of type~1 and type~2 particles show clearly distinct trends and features, indicating the two types correspond to early and late stages of incipient soot. The internal structure of the incipient particles from this investigation shows no direct or obvious sensitivity to temperature \editsf{for temperature ranging from 1350 to 1800 K}.

Other conclusions drawn from the study are
\begin{enumerate}
	\item Incipient soot particles comprise both cyclic and non-cyclic structures.
	\item The core of type~1 particles is dominated by non-cyclic structures while the core of type~2 particles is dominated by cyclic or ring structures.
	\item The internal distribution of ring and non-ring structures indicates the presence of a dense core region and a less dense shell region in type~2 incipient particles.
	\item The core comprises an interconnected cross-linked network of cyclic molecules while the shell region contains a sheet-like organization of cyclic molecules.
    \item The core of type~2 particles extend up to approximately $0.5-0.6R_g$, followed by the shell region beyond $0.6R_g$. \item The core and core-shell demarcation is not developed in type~1 particles.
\end{enumerate}

\section{Acknowledgments}
The research benefited from computational resources provided through the NCMAS, supported by the Australian Government, The University of Melbourne’s Research Computing Services and the Petascale Campus Initiative.  K.M.M. and S.P.R. acknowledge funding support from the National Science Foundation as some of this material is based upon work supported by the National Science Foundation under Grant No. 2144290. 

\bibliographystyle{elsarticle-num}
\bibliography{internal}

\appendix
\section{Symbols and nomenclature}
\noindent
$a$: A parameter in Eqn~\ref{e:bulk-density}\\
$A$: Surface area of a particle (\AA$^2$)\\
$c$: A parameter in Eqn~\ref{e:bulk-density}\\
$D_f$: Atomic fractal dimension of a particle \\
$m_{p,i}$: Mass of $i^\textrm{th}$ particle (kg)\\
$M$: Molar mass of a particle (kg/kmol)\\
$M_p$: Mass of a particle (kg)\\
$\ncring$: Number of cyclic carbon atoms at a specific location\\
$\ncnonring$: Number of non-cyclic carbon atoms at a specific location\\
$N$: Total number of atoms in the entire particle\\
$\Nring$: Number of rings in the entire particle\\
$\Nfive$: Number of 5-membered rings in the entire particle\\
$\Nsix$: Number of 6-membered rings in the entire particle\\
$\Nseven$: Number of 7-membered rings in the entire particle\\
$\Nc$: Number of carbon atoms in the entire particle\\
$\Nh$: Number of hydrogen atoms in the entire particle\\
$\Ncring$: Number of cyclic carbon atoms in the entire particle\\
$\Ncnonring$ Number of non-cyclic carbon atoms in the entire particle\\
$r$: Local radius (\AA)\\
$R_{eq}$: Volume equivalent radius (\AA)\\
$R_g$: Radius of gyration (\AA)\\
$\rho_s$: Simulated density of a particle (kg/m$^3$)\\
$\rho_e$: Empirical density of a particle (kg/m$^3$)\\
$\varrho$: Local (actual) density (kg/m$^3$)\\
$T$: Temperature (K)\\
$\chrl$: Local C/H ratio\\
$\chr$: C/H ratio of the entire particle\\
$V$: Volume of a particle (\AA$^3$)\\
$w_C$: Mass of a carbon atom (kg)\\
$w_H$: Mass of a hydrogen atom (kg)\\
\\
Superscripts:\\
$*$: Denotes normalized value\\
$\prime\prime$: Denotes per unit area value

\section{Expressions for physical properties of soot particles}
\label{app_eqn}
The trajectory files obtained from RMD simulations contain coordinates of each atom with reference to a global reference frame. This coordinate information along with the mass of each atom is used to calculate the coordinate of the center of mass of each particle. The mass of a particle $M_p$ is calculated by summing up the mass of the atoms in the cluster. The volume ($V$) is calculated using MSMS~\cite{Sanner1996Mar} with a pore size of 1.5~\AA.
The volume equivalent radius of a particle with volume $V$ is calculated via Eqn.~\ref{eq:R_eq}.
\begin{equation} \label{eq:R_eq}
	R_{eq} = \left(\frac{3V}{4\pi}\right)^{\nicefrac{1}{3}}
\end{equation}
The radius of gyration ($R_g$) is calculated following the standard definition using Eqn.~\ref{e:rg}.
\begin{equation}
	\label{e:rg}
	R_g = \sqrt{\frac{\sum_{i=1}^N m_{p,i} r_i^2}{\sum_{i=1}^N m_{p,i}}},
\end{equation}
where $r_i$ is the distance of the $i^{th}$ atom from the center of mass,
$m_{p,i}$ is the mass of individual atoms, and
$N$ is the total number of atoms in the cluster.

The simulated density ($\rho_{s}$) is calculated using the particle mass ($M_p$) and volume ($V$) of the incipient particle using
Eqn.~\ref{e:actual-denstiy}.
\begin{equation}
	\rho_{s} = \frac{M_p}{V}\label{e:actual-denstiy}
\end{equation}
Empirical (bulk) density \cite{Johansson2017Dec,DeCarlo2004Jan} of an incipient particle is calculated using Eqn.~\ref{e:bulk-density}.
\begin{eqnarray}
	\rho_{e} &= (0.260884 a^2c)^{-1}\left(\frac{w_C \chr +w_H}{\chr +1}\right),    \label{e:bulk-density}
\end{eqnarray}
where $w_C$ and $w_H$ are the molar masses of a carbon and hydrogen atoms, $a$ is the length of the graphite unite cell in
the basal plane, $c$ is the interlayer spacing in Angstroms, and $\chr$ represents the carbon to hydrogen ratio of the
cluster. More details can be found in \cite{Johansson2017Dec,DeCarlo2004Jan}.

The atomic fractal dimension ($D_f$), following the approach used in~\cite{Sharma2021Aug}, is calculated using the sandbox method \cite{Theiler1990Jun,Forrest1979May} using Eqn.~\ref{e:df}.
\begin{equation}
	\label{e:df}
	D_f = \frac{\log M_p(r)}{\log r},
\end{equation}
where $M_p(r)$ is the mass of atoms in the cluster as a function of radial distance from the center of mass. Please note that this ``atomic'' fractal dimension is for a single incipient particle and is different from the traditional fractal dimension used in aggregate characterization~\cite{Sharma2021Aug}.

\section{Physicochemical data used and analyzed in this study} 
\label{app_sample}

\subsection{Feature set}
\label{app_feature}
The feature set  used in this study for each particle includes the following
\begin{enumerate}
	\item Temperature ($T$)
	\item Number of carbon atoms ($N_C$)
	\item Number of hydrogen atoms ($N_H$)
	\item Number of atoms ($N$)
	\item Molar mass ($M$)
	\item C/H ratio ($\chr$)
	\item Radius of gyration ($R_g$)
	\item Atomic fractal dimension ($D_f$)
	\item Simulated density ($\rho_s$)
	\item Empirical density ($\rho_e$) (also referred to as the bulk density in literature)
	\item Total number of cyclic structures ($\Nring$)
	\item Fraction of cyclic carbon atoms ($\nicefrac{\Ncring}{\Nc}$)
	\item Fraction of 5-member rings ($\nicefrac{\Nfive}{\Nring}$)
	\item Fraction of 6-member rings ($\nicefrac{\Nsix}{\Nring}$)
	\item Fraction of 7-member rings ($\nicefrac{\Nseven}{\Nring}$)
	\item Surface area ($A$)
	\item Volume ($V$)
	\item Area to volume ratio ($\nicefrac{A}{V}$)
\end{enumerate}

\subsection{Sample data}
\label{app_sample-part}

\begin{figure} [!htbp]
	\centering
	\includegraphics[width=1\linewidth, height=3.25in]{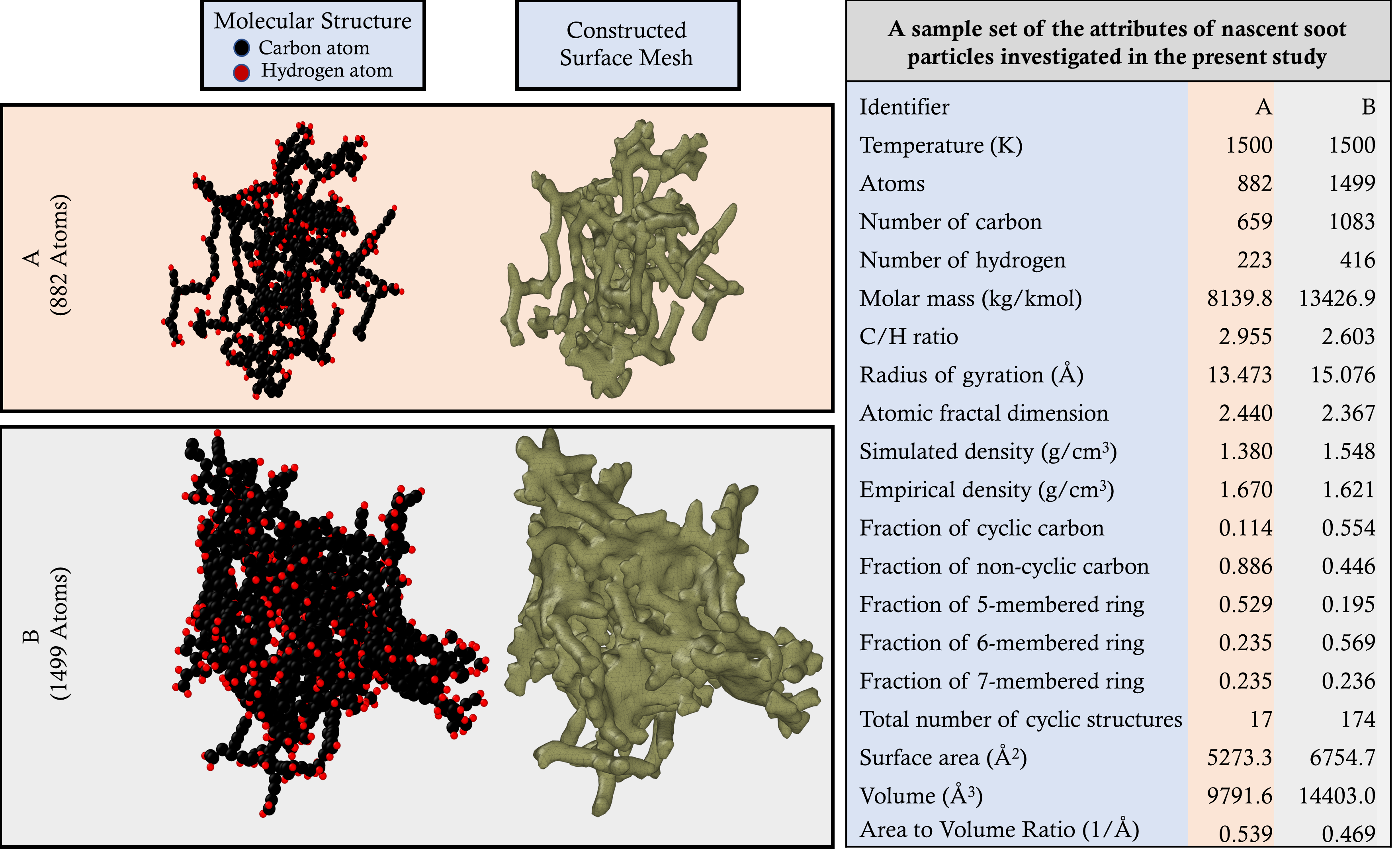}
	\caption{Two sample soot particles and their attributes investigated in this study}
	\label{f:sample}
\end{figure}

Figure~\ref{f:sample} shows two sample soot clusters and their properties as examples. These clusters, labeled as \verb\A\ and \verb\B\, were extracted from a simulation at 1500 K at two different times. The left side shows the molecular structure of the particle and the three-dimensional volumetric representation by constructing a surface mesh using \texttt{OVITO} \cite{Stukowski2009Dec} (this is what the incipient particles would actually look like). The physicochemical properties of these particles as analyzed in this work for classification via machine-learning are tabulated on the right side of the figure.
\end{document}